\documentclass[aps,a4paper,superscriptaddress,showpacs,preprintnumbers,amsmath,amssymb]{revtex4}

\usepackage{ulem}

\usepackage{psfrag} \usepackage{graphicx} \usepackage{dcolumn}
\usepackage{color} \usepackage{latexsym,amsfonts} \usepackage{bm}
\usepackage{amssymb}
\begin{document}

\title{Tests of the Standard Model\\ in Neutron Beta Decay with
  Polarized Neutron and Electron\\ and Unpolarized Proton}

\author{A. N. Ivanov}\email{ivanov@kph.tuwien.ac.at}
\affiliation{Atominstitut, Technische Universit\"at Wien, Stadionallee
  2, A-1020 Wien, Austria}
\author{R.~H\"ollwieser}\email{roman.hoellwieser@gmail.com}\affiliation{Atominstitut,
  Technische Universit\"at Wien, Stadionallee 2, A-1020 Wien,
  Austria}\affiliation{Department of Physics, Bergische Universit\"at
  Wuppertal, Gaussstr. 20, D-42119 Wuppertal, Germany}
\author{N. I. Troitskaya}\email{natroitskaya@yandex.ru}
\affiliation{Atominstitut, Technische Universit\"at Wien, Stadionallee
  2, A-1020 Wien, Austria}
\author{M. Wellenzohn}\email{max.wellenzohn@gmail.com}
\affiliation{Atominstitut, Technische Universit\"at Wien, Stadionallee
  2, A-1020 Wien, Austria} \affiliation{FH Campus Wien, University of
  Applied Sciences, Favoritenstra\ss e 226, 1100 Wien, Austria}
\author{Ya. A. Berdnikov}\email{berdnikov@spbstu.ru}\affiliation{Peter
  the Great St. Petersburg Polytechnic University, Polytechnicheskaya
  29, 195251, Russian Federation}

\date{\today}

\begin{abstract}
We analyse the electron--energy and angular distribution of the
neutron $\beta^-$--decay with polarized neutron and electron and
unpolarized proton, calculated in Phys. Rev. C {\bf 95}, 055502 (2017)
within the Standard Model (SM), by taking into account the
contributions of interactions beyond the SM. After the absorption of
vector and axial vector contributions by the axial coupling constant
and Cabibbo--Kobayashi--Maskawa (CKM) matrix element (Bhattacharya
{\it et al.}, Phys. Rev. D {\bf 85}, 054512 (2012) and so on) these
are the contributions of scalar and tensor interactions only. The
neutron lifetime, correlation coefficients and their averaged values,
and asymmetries of the neutron $\beta^-$--decay with polarized neutron
and electron are adapted to the analysis of experimental data on
searches of contributions of interactions beyond the SM. Using the
obtained results we propose some estimates of the values of the scalar
and tensor coupling constants of interactions beyond the SM. We use
the estimate of the Fierz interference term $b = - 0.0028 \pm
0.0026$ by Hardy and Towner (Phys. Rev. C {\bf 91}, 025501 (2015)),
the neutron lifetime $\tau_n = 880.2(1.0)\,{\rm s}$ (Particle Data
Group, Chin. Phys. C {\bf 40}, 100001 (2016)) and the experimental
data $N_{\exp} = 0.067 \pm 0.011_{\rm stat.} \pm 0.004_{\rm syst.}$
for the averaged value of the correlation coefficient of the
neutron--electron spin--spin correlations, measured by Kozela {\it et
  al.} (Phys. Ref. C {\bf 85}, 045501 (2012)). The contributions of
$G$--odd correlations are calculated and found at the level of
$10^{-5}$ in agreement with the results obtained by Gardner and
Plaster (Phys. Rev. C {\bf 87}, 065504 (2013)).
\end{abstract} 

\maketitle

\section{Introduction}
\label{sec:introduction}

Recently \cite{Ivanov2017b} we have calculated in the Standard Model
(SM) the electron--energy and angular distribution of the neutron
$\beta^-$--decay with polarized neutron and electron and unpolarized
proton by taking into account the contributions of the weak magnetism
and proton recoil of order $O(E_e/M)$, where $M$ is an averaged
nucleon mass and $E_e$ is the electron energy, and the radiative
corrections of order $O(\alpha/\pi)$, where $\alpha$ is the
fine--structure constant \cite{PDG2016}. These contributions define a
complete set of corrections of order $10^{-3}$ to the correlation
coefficients of the neutron $\beta^-$--decay with polarized neutron
and electron and unpolarized proton. The obtained results together
with Wilkinson's corrections of order $10^{-5}$ \cite{Wilkinson1982},
which we have also adapted to the correlation coefficients of the
neutron $\beta^-$--decay under consideration \cite{Ivanov2017b}, may
provide a robust SM theoretical background for the analysis of
experimental data on the searches of contributions of interactions
beyond the SM at the level of $10^{-4}$ \cite{Dubbers2008} or even
better \cite{Abele2016} (see also \cite{Ivanov2017b,Ivanov2013}) if
they are supplemented by a complete set of corrections of order
$10^{-5}$. This set of corrections is caused by the weak magnetism and
proton recoil of order $O(E^2_e/M^2)$, calculated to
next--to--next--to--leading order in the large nucleon mass expansion,
the radiative corrections of order $O(\alpha E_e/M)$, calculated to
next--to--leading order in the large nucleon mass expansion, and the
radiative corrections of order $O(\alpha^2/\pi^2)$, calculated to
leading order in the large nucleon mass expansion
\cite{Ivanov2017a,Ivanov2017c}.  The first steps towards the
experimental searches of contributions of interactions beyond the SM
in the neutron $\beta^-$--decay with polarized neutron and electron
and unpolarized proton have been done by Kozela {\it et al.}
\cite{Kozela2009,Kozela2012}.

The paper is organized as follows. In section \ref{sec:sm} we give the
electron--energy spectrum and angular distribution of the neutron
$\beta^-$--decay with polarized neutron and electron and unpolarized
proton, which has been calculated within the SM in \cite{Ivanov2017b}.
The correlation coefficients $A_W(E_e)$, $G(E_e)$, $N(E_e)$,
$Q_e(E_e)$ and $R(E_e)$ (see Eq.(\ref{eq:1})) are calculated at the
level of $10^{-3}$ by taking into account the contributions of the
weak magnetism and proton recoil to next--to--leading order in the
large proton mass expansion and radiative corrections of order
$O(\alpha/\pi)$, calculated to leading order in the large proton mass
expansion \cite{Ivanov2017b,Ivanov2013}. In section \ref{sec:bsm} we
calculate the contributions of interactions beyond the SM to the
correlation coefficients $A_W(E_e)$, $G(E_e)$, $N(E_e)$, $Q_e(E_e)$
and $R(E_e)$ , calculated to leading order in the large nucleon mass
expansion, and arrive at the correlation coefficients $A_{W, \rm
  eff}(E_e)$ $G_{\rm eff}(E_e)$, $N_{\rm eff}(E_e)$, $Q_{e,\rm
  eff}(E_e)$ and $R_{\rm eff}(E_e)$ \cite{Lee1956}--\cite{Gardner2013}
(see also \cite{Ivanov2013})). In the linear approximation for vector
and axial--vector interactions beyond the SM the obtained
contributions are defined by scalar and tensor nucleon--lepton
four--fermion couplings beyond the SM only in agreement with
\cite{Cirigliano2010,Bhattacharya2012,Cirigliano2013,Cirigliano2013a,Gardner2001,Gardner2013}
(see also \cite{Ivanov2013}). A possible dominant role of scalar and
tensor interactions beyond the SM has been also discussed by Jackson
{\it et al.}  \cite{Jackson1957}. In section \ref{sec:asymmetry} we
give the neutron lifetime, correlation coefficients $A_{W, \rm
  eff}(E_e)$, $G_{\rm eff}(E_e)$, $N_{\rm eff}(E_e)$, $Q_{e,\rm
  eff}(E_e)$ and $R_{\rm eff}(E_e)$ and asymmetries of the neutron
$\beta^-$--decay with polarized neutron and electron in the form
suitable for the analysis of experimental data of experiments on
searches of interactions beyond the SM, including the complete set of
the SM corrections of order $10^{-3}$, Wilkinson's corrections of
order $10^{-5}$ \cite{Ivanov2017b} and contributions of interactions
beyond the SM. The contributions of interactions beyond the SM agree
well with the results obtained by Jackson {\it et al.}
\cite{Jackson1957,Jackson1957a} and Severijns {\it et al.}
\cite{Severijns2006} up to redefinition of the metric and
normalization. In section \ref{sec:gparity} we calculation the
$G$--odd corrections to the neutron lifetime and correlation
coefficients of the neutron $\beta^-$--decay with polarized neutron
and electron and unpolarized proton. We estimate these corrections at
the level of $10^{-5}$ and even smaller in agreement with the results
obtained by Gardner and Plaster \cite{Gardner2013}.  In section
\ref{sec:conclusion} we discuss the obtained results and propose some
estimates of the values of scalar and tensor coupling constants of
interactions beyond the SM. We follow Severijns {\it et al.}
\cite{Severijns2006} and use for simplicity a real coupling constant
approximation and nucleon--lepton four--fermion couplings with
left--handed neutrinos only. The obtained results are adduced in Table
I and Table II. For the analysis of the experimental data of
experiments on the searches of contributions of interactions beyond
the SM at the level of $10^{-4}$ and even better \cite{Abele2016} we
argue an important role of the theoretical background with the SM
corrections of order $10^{-5}$ including Wilkinson's corrections
\cite{Ivanov2017b} and corrections, caused by the weak magnetism and
proton recoil of order $O(E^2_e/M^2)$, the radiative corrections of
order $O(\alpha E_e/M)$, and the radiative corrections of order
$O(\alpha^2/\pi^2)$ \cite{Ivanov2017a,Ivanov2017c}.

\section{Electron--energy and angular distribution in the SM}
\label{sec:sm}

The electron--energy and angular distribution of the neutron
$\beta^-$--decay with polarized neutron and electron, 
  introduced for the first time by Jackson {\it et al.}
  \cite{Jackson1957,Jackson1957a} but taken in notations
  \cite{Ivanov2017b}, is given by
\begin{eqnarray}\label{eq:1}
\hspace{-0.3in}\frac{d^3 \lambda_n(E_e,\vec{k}_e,
  \vec{\xi}_n,\vec{\xi}_e)}{dE_e d\Omega_e} &=& (1 + 3
\lambda^2)\,\frac{G^2_F|V_{ud}|^2}{8\pi^4} \,(E_0 - E_e)^2 \sqrt{E^2_e
  - m^2_e}\, E_e\,F(E_e, Z = 1)\,\zeta(E_e)\,\Big\{1 +
A_W(E_e)\,\frac{\vec{\xi}_n\cdot \vec{k}_e}{E_e}\nonumber\\
\hspace{-0.3in}&& + G(E_e)\,\frac{\vec{\xi}_e \cdot \vec{k}_e}{E_e} +
N(E_e)\,\vec{\xi}_n\cdot \vec{\xi}_e +
Q_e(E_e)\,\frac{(\vec{\xi}_n\cdot \vec{k}_e)( \vec{k}_e\cdot
  \vec{\xi}_e)}{E_e (E_e + m_e)} +
R(E_e)\,\frac{\vec{\xi}_n\cdot(\vec{k}_e \times
  \vec{\xi}_e)}{E_e}\Big\},
\end{eqnarray}
where $G_F = 1.1664\times 10^{-11}\,{\rm MeV}^{-2}$ is the Fermi weak
constant, $V_{ud} = 0.97417(21)$ is the Cabibbo-Kobayashi--Maskawa
(CKM) matrix element \cite{PDG2016}, extracted from the $0^+ \to 0^+$
transitions, $\lambda = - 1.2750(9)$ is the axial coupling constant,
which is real \cite{Ivanov2013}, $E_0 = (m^2_n - m^2_p + m^2_e)/2 m_n
= 1.2927\,{\rm MeV}$ is the end--point energy of the electron--energy
spectrum, calculated for $m_n = 939.5654\,{\rm MeV}$, and $m_p =
938.2721\,{\rm MeV}$ and $m_e = 0.5110\,{\rm MeV}$ \cite{PDG2016},
$\vec{\xi}_n$ and $\vec{\xi}_e$ are unit polarization vectors of the
neutron and electron, respectively, $F(E_e, Z = 1)$ is the
relativistic Fermi function
\cite{Jackson1957a,Blatt1952,Jackson1958,Konopinski1966}
\begin{eqnarray}\label{eq:2}
\hspace{-0.3in}F(E_e, Z = 1 ) = \Big(1 +
\frac{1}{2}\gamma\Big)\,\frac{4(2 r_pm_e\beta)^{2\gamma}}{\Gamma^2(3 +
  2\gamma)}\,\frac{\displaystyle e^{\,\pi
 \alpha/\beta}}{(1 - \beta^2)^{\gamma}}\,\Big|\Gamma\Big(1 + \gamma +
 i\,\frac{\alpha }{\beta}\Big)\Big|^2,
\end{eqnarray}
where $\beta = k_e/E_e = \sqrt{E^2_e - m^2_e}/E_e$ is the electron
velocity, $\gamma = \sqrt{1 - \alpha^2} - 1$, $r_p$ is the electric
radius of the proton.  In the numerical calculations we will use $r_p
= 0.841\,{\rm fm}$ \cite{Pohl2010} used in \cite{Ivanov2017a}, which
is smaller than $r_p = 0.875\,{\rm fm}$ reported in \cite{LEP2} and
used in \cite{Ivanov2017b}. The Fermi function Eq.(\ref{eq:2})
describes the contribution of the electron--proton final--state
Coulomb interaction. The analysis of different
  approximations of the Fermi function Eq.(\ref{eq:2}) has been
  carried out by Wilkinson \cite{Wilkinson1982} (see also
  \cite{Ivanov2017b}). In the SM the correlation coefficients of the
electron--energy and angular distribution Eq.(\ref{eq:1}) we calculate
with the Hamiltonian of $V-A$ weak interactions and the weak magnetism
\cite{Ivanov2013}
\begin{eqnarray}\label{eq:3}
\hspace{-0.3in}{\cal H}_W(x) =
\frac{G_F}{\sqrt{2}}\,V_{ud}\,\Big\{[\bar{\psi}_p(x)\gamma_{\mu}(1+
  \lambda \gamma^5)\psi_n(x)] + \frac{\kappa}{2 M}
\partial^{\nu}[\bar{\psi}_p(x)\sigma_{\mu\nu}\psi_n(x)]\Big\}
        [\bar{\psi}_e(x)\gamma^{\mu}(1 - \gamma^5)\psi_{\nu_e}(x)],
\end{eqnarray}
where $\psi_p(x)$, $\psi_n(x)$, $\psi_e(x)$ and $\psi_{\nu_e}(x)$ are
the field operators of the proton, neutron, electron and
anti-neutrino, respectively, $\gamma^{\mu}$, $\sigma^{\mu\nu} =
\frac{i}{2}(\gamma^{\mu}\gamma^{\nu} - \gamma^{\nu}\gamma^{\mu})$ and
$\gamma^5$ are the Dirac matrices; $\kappa = \kappa_p - \kappa_n =
3.7058$ is the isovector anomalous magnetic moment of the nucleon,
defined by the anomalous magnetic moments of the proton $\kappa_p =
1.7928$ and neutron $\kappa_n = - 1.9130$ and measured in nuclear
magneton \cite{PDG2016}, and $M = (m_n + m_p)/2$ is the average
nucleon mass.  The correlation coefficients $\zeta(E_e)$ and
$A_W(E_e)$ have been calculated in \cite{Ivanov2013}. They read
\begin{eqnarray}\label{eq:4}
\hspace{-0.3in}\zeta(E_e) &=&\Big(1 +
\frac{\alpha}{\pi}\,g_n(E_e)\Big) + \frac{1}{M}\,\frac{1}{1 +
  3\lambda^2}\,\Big[- 2\,\Big(\lambda^2 - (\kappa +
  1)\,\lambda\Big)\,E_0 + \Big(10 \lambda^2 - 4(\kappa + 1)\,\lambda +
  2\Big)\,E_e\nonumber\\
 \hspace{-0.3in}&-& 2 \,\Big(\lambda^2 - (\kappa +
  1)\,\lambda\Big)\,\frac{m^2_e}{E_e}\Big],\nonumber\\
\hspace{-0.3in}\zeta(E_e)\,A_W(E_e) &=& \zeta(E_e)\,\Big(A(E_e) +
\frac{1}{3}\,Q_n(E_e)\Big) = A_0\,\Big(1 +
\frac{\alpha}{\pi}\,g_n(E_e) +
\frac{\alpha}{\pi}\,f_n(E_e)\Big) +  \frac{1}{M}\,\frac{1}{1 +
   3\lambda^2}\nonumber\\
\hspace{-0.3in}&&\times\,\Big[\Big(\frac{4}{3}\,\lambda^2 -
  \Big(\frac{4}{3}\kappa + \frac{2}{3}\Big)\,\lambda -
  \frac{2}{3}(\kappa + 1)\Big)\,E_0 - \Big(\frac{22}{3}\lambda^2 -
  \Big(\frac{10}{3}\kappa - \frac{4}{3}\Big)\,\lambda -
  \frac{2}{3}(\kappa + 1)\Big)\,E_e\Big],
\end{eqnarray}
where the correlation coefficients $A(E_e)$ and $Q_n(E_e)$ are given
in \cite{Ivanov2013} (see also \cite{Gudkov2006}). The correlation
coefficient $A_W(E_e)$ without the contribution of the radiative
corrections, defined by the function $f_n(E_e)$, has been calculated
by Wilkinson \cite{Wilkinson1982}. We would like to remind that for
the first time the calculation of the corrections to order $O(E_e/M)$
to the correlation coefficients of the neutron $\beta^-$--decay with
polarized neutron and unpolarized proton and electron, caused by the
weak magnetism and proton recoil, has been carried out by Bilen'kii
{\it et al.}  \cite{Bilenky1959,Bilenky1960}.  The radiative
corrections $g_n(E_e)$ and $f_n(E_e)$ have been calculated by Sirlin
\cite{Sirlin1967} and Shann \cite{Shann1971}, respectively (for
details of these calculations one may consult \cite{Gudkov2006} and
\cite{Ivanov2013}).  The calculation of the contributions of the
$W$--boson and $Z$--boson exchanges and the QCD corrections to the
function $g_n(E_e)$ have been performed by Czarnecki {\it et al.}
\cite{Sirlin2004}.  The other correlation coefficients in the
electron--energy and angular distribution Eq.(\ref{eq:1}) have been
calculated in \cite{Ivanov2017b}. They are equal to
\begin{eqnarray}\label{eq:5}
\hspace{-0.3in}G(E_e) &=& - \Big(1 +
\frac{\alpha}{\pi}\,f_n(E_e)\Big)\,\Big(1 + \frac{1}{M}\,\frac{1}{1 +
  3\lambda^2}\,\Big(2 \lambda^2 - 2 (\kappa +
1)\,\lambda\Big)\,\frac{m^2_e}{E_e}\Big),\nonumber\\
\hspace{-0.3in}N(E_e) &=& + \Big(1 +
\frac{\alpha}{\pi}\,h^{(1)}_n(E_e)\Big)\,\frac{m_e}{E_e}\,\Big\{ - A_0
+ \frac{1}{M}\,\frac{1}{1 + 3
  \lambda^2}\,\Big[\Big(\frac{16}{3}\,\lambda^2 -
  \Big(\frac{4}{3}\kappa - \frac{16}{3}\Big)\,\lambda -
  \frac{2}{3}(\kappa + 1)\Big) E_e\nonumber\\
\hspace{-0.3in}&& -\Big(\frac{4}{3}\,\lambda^2 -
\Big(\frac{4}{3}\kappa - \frac{1}{3}\Big)\,\lambda -
\frac{2}{3}(\kappa + 1)\Big) E_0\Big] - \frac{1}{M}\, \frac{A_0}{1 + 3
  \lambda^2}\,\Big[- \Big(10 \lambda^2 - 4(\kappa + 1)\,\lambda +
  2\Big)\,E_e\nonumber\\
\hspace{-0.3in}&& + \Big(2 \lambda^2 - 2(\kappa +
1)\,\lambda\Big)\,\Big(E_0 +
\frac{m^2_e}{E_e}\Big)\Big]\Big\},\nonumber\\
\hspace{-0.3in}Q_e(E_e) &=& \,\Big(1 +
\frac{\alpha}{\pi}\,h^{(2)}_n(E_e)\Big)\,\Big\{ - A_0 +
\frac{1}{M}\,\frac{1}{1 + 3
  \lambda^2}\,\Big[\Big(\frac{22}{3}\,\lambda^2 - \Big(\frac{10}{3}
  \kappa - \frac{10}{3}\Big)\,\lambda - \frac{2}{3}(\kappa + 1)\Big)
  E_e\nonumber\\
\hspace{-0.3in}&&- \Big(\frac{4}{3}\,\lambda^2 - \Big(\frac{4}{3}
\kappa - \frac{1}{3}\Big)\,\lambda - \frac{2}{3}(\kappa + 1)\Big) E_0
+ \Big(2 \lambda^2 - 2 (\kappa + 1)\, \lambda\Big) m_e\Big] -
\frac{1}{M}\, \frac{A_0}{1 + 3 \lambda^2}\nonumber\\
\hspace{-0.3in}&&\,\Big[- \Big(10 \lambda^2 - 4(\kappa + 1)\,\lambda +
  2\Big)\,E_e + \Big(2 \lambda^2 - 2(\kappa +
  1)\,\lambda\Big)\,\Big(E_0 +
  \frac{m^2_e}{E_e}\Big)\Big]\Big\},\nonumber\\
\hspace{-0.3in}R(E_e) &=& - \alpha\,\frac{m_e}{k_e}\,A_0\quad,\quad
A_0 = - 2\,\frac{\lambda(1 + \lambda)}{1 + 3\lambda^2},
\end{eqnarray}
where the terms of order $ (\alpha/\pi)(E_e/M) < 3\times 10^{-6}$ are
neglected. The correlation coefficients Eq.(\ref{eq:5}) are defined at
the level of $10^{-3}$ of a complete set of contributions, caused by
the weak magnetism and proton recoil of order $O(E_e/M)$ and radiative
corrections of order $O(\alpha/\pi)$ \cite{Ivanov2017b}. The functions
$h^{(1)}_n(E_e)$ and $h^{(2)}_n(E_e)$, defining the radiative
corrections to the correlation coefficients of $N(E_e)$ and
$Q_e(E_e)$, are calculated for the first time in \cite{Ivanov2017b}.

\section{Electron--energy and angular distribution beyond the SM}
\label{sec:bsm}

For the calculation of contributions of interactions beyond the SM we
use the effective low--energy Hamiltonian of weak nucleon--lepton
four--fermion local interactions, taking into account all
phenomenological couplings beyond the SM
\cite{Lee1957}--\cite{Severijns2006}. In notations \cite{Ivanov2013}
such a Hamiltonian takes the form
\begin{eqnarray*}
{\cal H}_W(x) &=&
\frac{G_F}{\sqrt{2}}\,V_{ud}\Big\{[\bar{\psi}_p(x)\gamma_{\mu}\psi_n(x)]
     [\bar{\psi}_e(x)\gamma^{\mu}(C_V + \bar{C}_V
       \gamma^5)\psi_{\nu_e}(x)] +
     [\bar{\psi}_p(x)\gamma_{\mu}\gamma^5\psi_n(x)][\bar{\psi}_e(x)
       \gamma^{\mu}(\bar{C}_A + C_A \gamma^5)\psi_{\nu_e}(x)]
     \nonumber\\ &+& [\bar{\psi}_p(x)\psi_n(x)][\bar{\psi}_e(x)(C_S +
       \bar{C}_S \gamma^5)\psi_{\nu_e}(x)] + [\bar{\psi}_p(x) \gamma
       ^5 \psi_n(x)][\bar{\psi}_e(x)(C_P + \bar{C}_P
       \gamma^5)\psi_{\nu_e}(x)]\nonumber\\ 
\end{eqnarray*}
\begin{eqnarray}\label{eq:6}
&+&\frac{1}{2}
                 [\bar{\psi}_p(x)\sigma^{\mu\nu}\gamma^5\psi_n(x)]
                 [\bar{\psi}_e(x)\sigma_{\mu\nu} (\bar{C}_T + C_T
                   \gamma^5)\psi_{\nu_e}(x) \Big\}.
\end{eqnarray}
This is the most general form of the effective low--energy weak
interactions, where the phenomenological coupling constants $C_i$ and
$\bar{C}_i$ for $i = V, A, S, P$ and $T$ can be induced by the
left--handed and right--handed hadronic and leptonic currents
\cite{Jackson1957}--\cite{Herczeg2001} and \cite{Severijns2006}. They
are related to the phenomenological coupling constants, analogous to
those which were introduced by Herczeg \cite{Herczeg2001}, as follows
\begin{eqnarray}\label{eq:7}
\hspace{-0.3in} C_V &=&1 + a^h_{LL} + a^h_{LR} + a^h_{RR} +
a^h_{RL}\quad,\quad \bar{C}_V = - 1 - a^h_{LL} -
a^h_{LR} + a^h_{RR} + a^h_{RL},\nonumber\\
\hspace{-0.3in} C_A &=& -\lambda + a^h_{LL} - a^h_{LR} + a^h_{RR} -
a^h_{RL}\quad,\quad \bar{C}_A = \lambda - a^h_{LL} + a^h_{LR} + a^h_{RR} -
a^h_{RL},\nonumber\\
\hspace{-0.3in}C_S &=& A^h_{LL} + A^h_{LR} +
A^h_{RR} + A^h_{RL} \quad,\quad \bar{C}_S =  - A^h_{LL} -
A^h_{LR} + A^h_{RR} + A^h_{RL},\nonumber\\
\hspace{-0.3in}C_P &=& - A^h_{LL} + A^h_{LR} + A^h_{RR} -
A^h_{RL}\quad,\quad \bar{C}_P = A^h_{LL} - A^h_{LR} + A^h_{RR} -
A^h_{RL},\nonumber\\
\hspace{-0.3in} C_T &=& 2( \alpha^h_{LL} + \alpha^h_{RR})\quad,\quad
\bar{C}_T =  2( - \alpha^h_{LL} + \alpha^h_{RR}),
\end{eqnarray}
where the index $h$ means that the phenomenological coupling constants
are introduced at the {\it hadronic} level but not at the {\it quark}
level as it has been done by Herczeg \cite{Herczeg2001}. In the SM the
phenomenological coupling constants $C_i$ and $\bar{C}_i$ for $i = V,
A, S, P$ and $T$ are equal to $C_S = \bar{C}_S = C_P = \bar{C}_P = C_T
= \bar{C}_T = 0$, $C_V = - \,\bar{C}_V = 1$ and $C_A = - \,\bar{C}_A =
- \lambda$ \cite{Ivanov2013}.  The phenomenological coupling constants
$a^h_{ij}$, $A^h_{ij}$ and $\alpha^h_{jj}$ for $i(j) = L$ or $R$ are
induced by interactions beyond the SM. The coupling
  constants in Eq.(\ref{eq:6}) are related to the coupling constants
  by Gudkov {\it et al.} \cite{Gudkov2006}) as follows
\begin{eqnarray}\label{eq:8}
\hspace{-0.3in}&& C_V = C_V\;,\; \bar{C}_V = C'_V\;,\; \bar{C}_A = -
C_A\;,\; C_A = -C'_A\;,\; C_S = C_S\;,\; \bar{C}_S = C'_S,\nonumber\\
\hspace{-0.3in}&& C_P = C'_P\;,\; \bar{C}_P = C_P\;,\; C_T = C_T\;,\;
\bar{C}_T = C'_T.
\end{eqnarray}
Thus, our definition of the coupling constants of interactions beyond
the SM, used in \cite{Ivanov2013,Ivanov2014}, differs from Gudkov's
definition only for the axial--vector coupling
constants. However, the contributions to the correlation coefficients,
obtained in the linear approximation with respect to deviations of the
vector and axial--vector coupling constants from their values in the
SM and expressed in terms of the scalar and tensor coupling constants,
are related by the redefinition $(C_j,\bar{C}_j) \to (C_j, C'_j)$ for
$j = S,T$ (see Eq.(\ref{eq:15}) and section \ref{sec:asymmetry}).

The structure of the phenomenological coupling constants
Eq.(\ref{eq:7}) agrees well with the coupling constants of
interactions beyond the SM used by Cirigliano {\it et al.}
\cite{Cirigliano2013} for consideration of the role of precision
measurements of beta decays and light meson semi--leptonic decays in
probing physics beyond the SM in the LHC era. For this aim, using an
effective field theory framework, all low-energy charged-current
processes within and beyond the SM were described, and theoretical
hadronic input which in these precision tests plays a crucial role in
setting the baseline for new physics searches was discussed.

The contribution of interactions beyond the SM, given by the
Hamiltonian of weak interactions Eq.(\ref{eq:6}), to the amplitude of
the neutron $\beta^-$--decay, calculated to leading order in the large
nucleon mass expansion, takes the form
\begin{eqnarray}\label{eq:9}
\hspace{-0.3in} M(n \to p e^- \bar{\nu}_e) &=& -\,2m_n\,
\frac{G_F}{\sqrt{2}}\,V_{ud}\,\Big\{[\varphi^{\dagger}_p\varphi_n]
     [\bar{u}_e \gamma^0(C_V + \bar{C}_V \gamma^5) v_{\bar{\nu}}] -
     [\varphi^{\dagger}_p\vec{\sigma}\,\varphi_n]\cdot [\bar{u}_e
       \vec{\gamma}\,(\bar{C}_A + C_A \gamma^5)
       v_{\bar{\nu}}]\nonumber\\
\hspace{-0.3in}&&+ [\varphi^{\dagger}_p\varphi_n][\bar{u}_e (C_S +
  \bar{C}_S \gamma^5) v_{\bar{\nu}}] +
       [\varphi^{\dagger}_p\vec{\sigma}\,\varphi_n]\cdot [\bar{u}_e
         \gamma^0 \vec{\gamma}\,(\bar{C}_T + C_T \gamma^5)
         v_{\bar{\nu}}]\Big\}.
\end{eqnarray}
The hermitian conjugate amplitude is
\begin{eqnarray}\label{eq:10}
\hspace{-0.3in} M^{\dagger}(n \to p e^- \bar{\nu}_e) &=& -\,2m_n\,
\frac{G_F}{\sqrt{2}}\,V^*_{ud}\Big\{[\varphi^{\dagger}_n\varphi_p]
     [\bar{v}_{\bar{\nu}} \gamma^0(C^*_V + \bar{C}^*_V \gamma^5) u_e]
     - [\varphi^{\dagger}_n\vec{\sigma}\,\varphi_p]\cdot
     [\bar{v}_{\bar{\nu}} \vec{\gamma}\,(\bar{C}^*_A + C^*_A \gamma^5)
       u_e] \nonumber\\
\hspace{-0.3in}&&+ [\varphi^{\dagger}_n\varphi_p][\bar{v}_{\bar{\nu}}
  (C^*_S - \bar{C}^*_S \gamma^5) u_e] -
       [\varphi^{\dagger}_n\vec{\sigma}\,\varphi_p]\cdot
       [\bar{v}_{\bar{\nu}} \gamma^0 \vec{\gamma}\,(\bar{C}^*_T -
         C^*_T \gamma^5) u_e]\Big\}.
\end{eqnarray}
The contributions of interactions with the strength, defined by the
phenomenological coupling constants $C_P$ and $\bar{C}_P$, may appear
only of order $O(C_P E_e/M)$ and $O(\bar{C}_P E_e/M)$ and can be
neglected to leading order in the large nucleon mass expansion. We
have also neglected the contributions of the neutron--proton mass
difference. The squared absolute value of the amplitude
Eq.(\ref{eq:9}), summed over polarizations of massive fermions, is
equal to
\begin{eqnarray*}
\hspace{-0.15in}&& \sum_{\rm pol.}|M(n \to p e^- \bar{\nu}_e)|^2 =
8m^2_n G^2_F|V_{ud}|^2 E_{\nu}E_e\Big\{\frac{1}{2}\Big(|C_V|^2 +
|\bar{C}_V|^2 + 3|C_A|^2 + 3|\bar{C}_A|^2 + |C_S|^2 + |\bar{C}_S|^2 +
3 |C_T|^2 + 3|\bar{C}_T|^2\Big)\nonumber\\
\hspace{-0.15in}&& + \frac{m_e}{E_e}\,{\rm Re}\Big(C_VC^*_S +
\bar{C}_V\bar{C}^*_S - 3 C_AC^*_T - 3 \bar{C}_A\bar{C}^*_T\Big) +
\frac{\vec{\xi}_n\cdot \vec{k}_e}{E_e}\,{\rm Re}\Big( 2C_A \bar{C}^*_A
- 2 C_T \bar{C}^*_T - C_V \bar{C}^*_A - \bar{C}_V C^*_A - C_S
\bar{C}^*_T - \bar{C}_S C^*_T\Big)\nonumber\\
\hspace{-0.15in}&& + \frac{\vec{k}_e\cdot \vec{\xi}_e}{E_e}\,{\rm
  Re}\Big(C_V\bar{C}^*_V + 3 C_A\bar{C}^*_A - C_S\bar{C}^*_S - 3
C_T\bar{C}^*_T \Big) + \vec{\xi}_n\cdot \vec{\xi}_e\,{\rm
  Re}\Big[\frac{m_e}{E_e}\,\Big(|C_A|^2 + |\bar{C}_A|^2 - C_VC^*_A -
  \bar{C}_V \bar{C}^*_A + |C_T|^2 + |\bar{C}_T|^2 \nonumber\\\hspace{-0.15in}
\end{eqnarray*}
\begin{eqnarray}\label{eq:11}
\hspace{-0.15in}&& + C_S C^*_T + \bar{C}_S \bar{C}^*_T \Big) + C_V
C^*_T + \bar{C}_V \bar{C}^*_T - C_A C^*_S - \bar{C}_A \bar{C}^*_S - 2
C_A C^*_T - 2 \bar{C}_A \bar{C}^*_T\Big] +
\frac{(\vec{\xi}_n\cdot \vec{k}_e)(\vec{k}_e\cdot \vec{\xi}_e)}{E_e
  (E_e + m_e)}\,{\rm Re}\Big(|C_A|^2 + |\bar{C}_A|^2\nonumber\\
\hspace{-0.15in}&& - C_V C^*_A - \bar{C}_V \bar{C}^*_A - C_V C^*_T -
\bar{C}_V \bar{C}^*_T + C_A C^*_S + \bar{C}_A \bar{C}^*_S + 2 C_A
C^*_T + 2 \bar{C}_A \bar{C}^*_T + |C_T|^2 + |\bar{C}_T|^2 + C_S C^*_T
+ \bar{C}_S \bar{C}^*_T\Big)\nonumber\\
\hspace{-0.15in}&&+ \frac{\vec{\xi}_n\cdot (\vec{k}_e \times
  \vec{\xi}_e)}{E_e}\,{\rm Im}\Big(C_V\bar{C}^*_T + \bar{C}_V C^*_T -
C_A\bar{C}^*_S - \bar{C}_A C^*_S - 2 C_A\bar{C}^*_T - 2 \bar{C}_A
C^*_T \Big)\Big\}.
\end{eqnarray}
In Eq.(\ref{eq:11}) the structure of the contributions of interactions
beyond the SM agrees well with the structure of the corresponding
expressions obtained by Jackson {\it et al.}
\cite{Jackson1957,Jackson1957a}. The first term on the second line of
Eq.(\ref{eq:11}) is the Fierz interference term. It appears as a
result of the calculation of the traces over the Dirac matrices on the
same footing as it appeared in the paper by Lee and Yang
\cite{Lee1956} (see the Appendix of Ref.\cite{Lee1956} and
\cite{Rose1955}). In the linear approximation for coupling constants
of vector and axial--vector interactions beyond the SM
\cite{Ivanov2013} we get
\begin{eqnarray}\label{eq:12}
\hspace{-0.3in}&& \sum_{\rm pol.}|M(n \to p e^- \bar{\nu}_e)|^2 =
8m^2_n G^2_F|V_{ud}|^2 E_{\nu}E_e\,(1 + 3 \lambda^2)\,\Big\{\Big[1 +
  \frac{1}{2}\,\frac{1}{1 + 3\lambda^2}\,( |C_S|^2 + |\bar{C}_S|^2 + 3
  |C_T|^2 + 3|\bar{C}_T|^2)\Big]\nonumber\\
\hspace{-0.3in}&& + \frac{m_e}{E_e}\,\frac{1}{1 + 3\lambda^2}\,{\rm
  Re}\Big((C_S - \bar{C}_S) + 3 \lambda\,(C_T -\bar{C}_T)\Big) +
\frac{\vec{\xi}_n\cdot \vec{k}_e}{E_e}\,\Big( A_0 - \frac{1}{1 + 3
  \lambda^2}\,{\rm Re}( C_S \bar{C}^*_T + \bar{C}_S C^*_T + 2 C_T
\bar{C}^*_T)\Big)\nonumber\\
\hspace{-0.3in}&& + \frac{\vec{k}_e\cdot \vec{\xi}_e}{E_e}\,\Big(- 1
- \frac{1}{1 + 3\lambda^2}\,{\rm Re}\Big(C_S\bar{C}^*_S + 3
C_T\bar{C}^*_T\Big) \Big) + \vec{\xi}_n\cdot \vec{\xi}_e\,
\Big[\frac{m_e}{E_e}\,\Big( - A_0 + \frac{1}{1 + 3\lambda^2}\,{\rm
    Re}\Big(C_S C^*_T + \bar{C}_S \bar{C}^*_T \nonumber\\
\hspace{-0.3in}&& + |C_T|^2 + |\bar{C}_T|^2\Big)\Big) + \frac{1}{1 +
  3\lambda^2}\,{\rm Re}\Big(\lambda(C_S - \bar{C}_S) + (1 +
2\lambda)(C_T - \bar{C}_T)\Big)\Big] + \frac{(\vec{\xi}_n\cdot
  \vec{k}_e)(\vec{k}_e\cdot \vec{\xi}_e)}{E_e (E_e + m_e)}\,\Big[- A_0
  + \frac{1}{1 + 3\lambda^2}\nonumber\\
\hspace{-0.3in}&& \times\,{\rm Re}\Big(-\lambda( C_S -\bar{C}_S) - (1
+ 2\lambda)(C_T - \bar{C}_T) + C_S C^*_T + \bar{C}_S \bar{C}^*_T +
|C_T|^2 + |\bar{C}_T|^2\Big)\Big]\nonumber\\
\hspace{-0.3in}&&+ \frac{\vec{\xi}_n\cdot (\vec{k}_e \times
  \vec{\xi}_e)}{E_e}\, \frac{1}{1 + 3\lambda^2}\,{\rm Im}\Big(\lambda
(C_S - \bar{C}_S) + (1 + 2\lambda)(C_T - \bar{C}_T) \Big)\Big\},
\end{eqnarray}
where we have replaced $C_j$ and $\bar{C}_j$ with $j = V,A$ by $C_V =
1 + \delta C_V$, $\bar{C}_V = - 1 + \delta \bar{C}_V$, $C_A = -
\lambda + \delta C_A$ and $\bar{C}_A = \lambda + \delta \bar{C}_A$
\cite{Gudkov2006} (see also \cite{Ivanov2013}) and neglected the
contributions of the products $(\delta C_j)^2$, $\delta C_j\delta
\bar{C}_j$, $\delta C_j C_k$ and $\delta \bar{C}_j C_k$ for $j = V,A$
and $k = S,T$. Following \cite{Bhattacharya2012}) (see also
\cite{Ivanov2013}) we have absorbed the contributions the vector and
axial vector interactions beyond the SM by the axial coupling constant
$\lambda$ and the CKM matrix element $V_{ud}$.

Thus, the electron--energy and angular distribution
  Eq.(\ref{eq:1}), taking into account the contributions of
  interactions beyond the SM, can be transcribed into the following
  standard form \cite{Jackson1957} (see also \cite{Gudkov2006} and
  \cite{Bhattacharya2012}--\cite{Gardner2013})
\begin{eqnarray}\label{eq:13}
&&\frac{d^3 \lambda_n(E_e,\vec{k}_e, \vec{\xi}_n,\vec{\xi}_e)}{dE_e
    d\Omega_e} = (1 + 3 \lambda^2)\,\frac{G^2_F|V_{ud}|^2}{8\pi^4}
  (E_0 - E_e)^2 \sqrt{E^2_e - m^2_e}\, E_e F(E_e, Z = 1) \zeta^{(\rm
    SM)}(E_e) \nonumber\\ &&\times \Big(1 + \zeta^{(\rm BSM)}(E_e)
  \Big)\Big\{1 + b\,\frac{m_e}{E_e} + A_{W,\rm eff}(E_e)\,
  \frac{\vec{\xi}_n\cdot \vec{k}_e}{E_e} + G_{\rm eff}(E_e)
  \frac{\vec{\xi}_e \cdot \vec{k}_e}{E_e} + N_{\rm eff}(E_e)
  \vec{\xi}_n\cdot \vec{\xi}_e \nonumber\\ &&+ Q_{e, \rm eff}(E_e)
  \frac{(\vec{\xi}_n\cdot \vec{k}_e)( \vec{k}_e\cdot \vec{\xi}_e)}{E_e
    (E_e + m_e)} + R_{\rm eff}(E_e) \frac{\vec{\xi}_n\cdot(\vec{k}_e
    \times \vec{\xi}_e)}{E_e}\Big\},
\end{eqnarray}
where the indices ``SM'' and ``BSM'' mean ``Standard Model'' and
``Beyond Standard Model'', respectively. The correlation coefficient
$\zeta^{(\rm SM)}(E_e)$ is given by Eq.(\ref{eq:4}), whereas the
analytical expressions for the correlation coefficient $\zeta^{(\rm
  BSM)}(E_e)$, are adduced in Eq.(\ref{eq:15}). Other correlation
coefficients are defined by
\begin{eqnarray}\label{eq:14}
b &=& \frac{b_F}{\displaystyle 1 + \zeta^{(\rm BSM)}(E_e)}\quad,\quad
A_{W, \rm eff}(E_e) = \frac{A^{(\rm SM)}_W(E_e) + A^{(\rm
    BSM)}_W(E_e)}{\displaystyle 1 + \zeta^{(\rm BSM)}(E_e)},
\nonumber\\ G_{\rm eff}(E_e) &=& \frac{G^{(\rm SM)}(E_e) + G^{(\rm
    BSM)}(E_e)}{\displaystyle 1 + \zeta^{(\rm BSM)}(E_e)}\quad,\quad
N_{\rm eff}(E_e) = \frac{N^{(\rm SM)}(E_e) + N^{(\rm
    BSM)}(E_e)}{\displaystyle 1 + \zeta^{(\rm
    BSM)}(E_e)},\nonumber\\ Q_{e,\rm eff}(E_e) &=& \frac{Q^{(\rm
    SM)}_e(E_e) + Q^{(\rm BSM)}_e(E_e)}{\displaystyle 1 + \zeta^{(\rm
    BSM)}(E_e)}\quad,\quad R_{\rm eff}(E_e) = \frac{R^{(\rm SM)}(E_e)
  + R^{(\rm BSM)}(E_e)}{\displaystyle 1 + \zeta^{(\rm BSM)}(E_e)},
\end{eqnarray}
where $b$ is the Fierz interference term.  The correlation
coefficients with index ``SM'' are given by Eqs.(\ref{eq:4}) and
(\ref{eq:5}). These expressions should be also supplemented by
Wilkinson's corrections of order $10^{-5}$ \cite{Wilkinson1982},
calculated for the neutron $\beta^-$--decay under consideration in
\cite{Ivanov2017b}. The correlation coefficients $b_F$, $b_E$ and
others with index ``BSM'' are given by
\begin{eqnarray}\label{eq:15}
b_F &=& \frac{1}{1 + 3\lambda^2}\,{\rm Re}((C_S - \bar{C}_S) + 3
\lambda\,(C_T -\bar{C}_T)),\nonumber\\ b_E &=& \frac{1}{1 +
  3\lambda^2}\,{\rm Re}(\lambda\, (C_S - \bar{C}_S) + (1 + 2\lambda)\,
(C_T -\bar{C}_T)),\nonumber\\
\hspace{-0.3in}\zeta^{(\rm BSM)}(E_e) &=& \frac{1}{2}\,\frac{1}{1 +
  3\lambda^2}\,( |C_S|^2 + |\bar{C}_S|^2 + 3 |C_T|^2 +
3|\bar{C}_T|^2),\nonumber\\ A^{(\rm BSM)}_W(E_e) &=&- \frac{1}{1 + 3
  \lambda^2}\,{\rm Re}( C_S \bar{C}^*_T + \bar{C}_S C^*_T + 2 C_T
\bar{C}^*_T),\nonumber\\ G^{(\rm
  BSM)}(E_e) &=& - \frac{1}{1 + 3\lambda^2}\,{\rm Re}(C_S\bar{C}^*_S +
3 C_T\bar{C}^*_T),\nonumber\\ N^{(\rm
  BSM)}(E_e)&=& \frac{m_e}{E_e}\, \frac{1}{1 + 3\lambda^2}\,{\rm
  Re}(C_S C^*_T + \bar{C}_S \bar{C}^*_T + |C_T|^2 + |\bar{C}_T|^2) +
b_E,\nonumber\\ Q^{(\rm BSM)}_e(E_e)&=&
\frac{1}{1 + 3\lambda^2}\,{\rm Re}(C_S C^*_T + \bar{C}_S \bar{C}^*_T +
|C_T|^2 + |\bar{C}_T|^2) - b_E,
\nonumber\\ R^{(\rm BSM)}(E_e)&=& \frac{1}{1 + 3\lambda^2}\,{\rm
  Im}(\lambda (C_S - \bar{C}_S) + (1 + 2 \lambda)(C_T - \bar{C}_T)).
\end{eqnarray}
The correlation coefficients in Eq.(\ref{eq:15}) can be redefined in
Gudkov's notation \cite{Gudkov2006}) by a replacement $(C_, \bar{C}_j)
\to (C_j, C'_j)$ for $j = S,T$ (see Eq.(\ref{eq:8})).  In
Eq.(\ref{eq:15}) the structure of the contributions of interactions
beyond the SM agrees well with the structure of corresponding
expressions taken in the linear approximation with respect to vector
and axial--vector interactions beyond the SM obtained by Jackson {\it
  et al.}  \cite{Jackson1957,Jackson1957a}. For the calculation of
Eq.(\ref{eq:13}) we have carried out the integration over the
directions of the antineutrino momentum. This gives the correlation
coefficient $A^{(\rm SM)}_W(E_e)$ equal to $A^{(\rm SM)}_W(E_e) =
A^{(\rm SM)}(E_e) + \frac{1}{3}\,Q^{(\rm SM)}_n(E_e)$
\cite{Wilkinson1982,Ivanov2013} (see also Eq.(\ref{eq:26})).


\section{Neutron lifetime, averaged values of correlation 
coefficients and asymmetries of neutron $\beta^-$--decay with
polarized neutron and electron}
\label{sec:asymmetry}

For the analysis of experimental data of experiments on the searches of
interactions beyond the SM in the neutron $\beta^-$--decay with
polarized neutron and electron we propose to use a complete set of
contributions of scalar and tensor interactions beyond the SM
including linear, crossing and quadratic terms, which are given in
Eq.(\ref{eq:15}).

\subsection{Neutron lifetime}

Having integrated the electron--energy and angular distribution
Eq.(\ref{eq:13}) with contributions of interactions beyond the SM we
get
\begin{eqnarray}\label{eq:16}
\tau^{-1}_n(\vec{\xi}_n,\vec{\xi}_e) = \tau^{-1}_n \Big(1 +
\frac{1}{2}\,\frac{1}{1 + 3\lambda^2}\,( |C_S|^2 + |\bar{C}_S|^2 + 3
|C_T|^2 + 3|\bar{C}_T|^2)+ \Big\langle \frac{m_e}{E_e}\Big\rangle_{\rm
  SM} b_F + \langle \bar{N}_{\rm eff}(E_e)\rangle \,\vec{\xi}_e\cdot
\vec{\xi}_e\Big).
\end{eqnarray}
Here we have denoted 
\begin{eqnarray}\label{eq:17}
\langle \bar{N}_{\rm eff}(E_e)\rangle = \Big\langle N^{(\rm SM)}(E_e)
+\frac{1}{3}\Big(1 - \frac{m_e}{E_e}\Big)\,Q^{(\rm
  SM)}_e(E_e)\Big\rangle_{\rm SM} + \langle \bar{N}^{(\rm
  BSM)}(E_e)\rangle_{\rm SM}.
\end{eqnarray}
For the calculation of the averaged value $\langle \bar{N}_{\rm
  eff}(E_e)\rangle$ we use the electron--energy density
\begin{eqnarray}\label{eq:18}
\rho_e(E_e) = \rho^{(\rm SM)}_e(E_e)\,\Big( 1 +
\frac{1}{2}\,\frac{1}{1 + 3\lambda^2}\,( |C_S|^2 + |\bar{C}_S|^2 + 3
|C_T|^2 + 3|\bar{C}_T|^2)\Big),
\end{eqnarray}
where the electron--energy density $\rho^{(\rm SM)}_e(E_e)$ is defined
by Eq.(D-59) of Ref.\cite{Ivanov2013}. The notation $\langle \ldots
\rangle_{\rm SM}$ means that the integration over the electron--energy
spectrum is carried out with the electron--energy density $\rho^{(\rm
  SM)}_e(E_e)$. Then, $\langle \bar{N}^{(\rm BSM)}(E_e)\rangle_{\rm
  SM}$ is equal to
\begin{eqnarray}\label{eq:19}
\langle \bar{N}^{(\rm BSM)}(E_e)\rangle_{\rm SM} = \Big(\frac{2}{3} +
\frac{1}{3}\,\Big\langle \frac{m_e}{E_e}\Big\rangle_{\rm SM}\Big)\,b_E
+ \Big(\frac{1}{3} + \frac{2}{3}\,\Big\langle
\frac{m_e}{E_e}\Big\rangle_{\rm SM}\Big)\,\frac{1}{1 +
  3\lambda^2}\,{\rm Re}\Big(C_S C^*_T + \bar{C}_S \bar{C}^*_T +
|C_T|^2 + |\bar{C}_T|^2\Big),
\end{eqnarray}
where $\langle m_e/E_e\rangle_{\rm SM} = 0.6556$ and $\tau_n =
879.6(1.1)\,{\rm s}$ \cite{Ivanov2013}. Recent analysis of the
experimental data on the neutron lifetime, carried out by Czarnecki
{\it et al.}  \cite{Sirlin2018}, has led to the {\it favoured} neutron
lifetime $\tau^{(\rm favoured)} = 879.4(6)\,{\rm s}$ and the {\it
  favoured} axial coupling constant $\lambda^{(\rm favoured)} = -
1.2755(11)$, which agree very well with $\tau_n = 879.6(1.1)\,{\rm s}$
and $\lambda = - 1.2750(9)$, respectively \cite{Ivanov2013}.

\subsection{Averaged values of correlations coefficients}

In terms of the correlation coefficients $b_F$ and $b_E$ the
phenomenological scalar and tensor coupling constants ${\rm Re}(C_S -
\bar{C}_S)$ and ${\rm Re}(C_T - \bar{C}_T)$ are defined by
\begin{eqnarray}\label{eq:20}
{\rm Re}(C_S - \bar{C}_S) &=& \frac{3\lambda^2 + 1}{3\lambda^2 -
  2\lambda -1}\,(- (1 + 2\lambda)\,b_F + 3\lambda\,b_E
),\nonumber\\ {\rm Re}(C_T - \bar{C}_T) &=& \frac{ 3\lambda^2 + 1}{
  3\lambda^2 - 2\lambda - 1}\,(\lambda\,b_F - b_E ).
\end{eqnarray}
The averaged values of the correlation coefficients $A_{W, \rm
  eff}(E_e)$, $G_{\rm eff}(E_e)$, $N_{\rm eff}(E_e)$, $Q_{e,\rm
  eff}(E_e)$ and $R_{\rm eff}(E_e)$, taking into account the
contributions of the SM and interactions beyond the SM, are given by
\begin{eqnarray}\label{eq:21}
\langle A_{W, \rm eff}(E_e)\rangle &=& \langle A^{(\rm
  SM)}_W(E_e)\rangle_{\rm SM} - \frac{1}{1 + 3 \lambda^2}\,{\rm Re}(
C_S \bar{C}^*_T + \bar{C}_S C^*_T + 2 C_T \bar{C}^*_T)
=\nonumber\\ &=& - 0.12121 - \frac{1}{1 + 3 \lambda^2}\,{\rm Re}( C_S
\bar{C}^*_T + \bar{C}_S C^*_T + 2 C_T \bar{C}^*_T),\nonumber\\ \langle
G_{\rm eff}(E_e)\rangle &=& \langle G^{(\rm SM)}(E_e)\rangle_{\rm SM}
- \frac{1}{1 + 3\lambda^2}\,{\rm Re}(C_S\bar{C}^*_S + 3
C_T\bar{C}^*_T) =\nonumber\\ &=&-1.00242 - \frac{1}{1 +
  3\lambda^2}\,{\rm Re}(C_S\bar{C}^*_S + 3
C_T\bar{C}^*_T),\nonumber\\ \langle N_{\rm eff}(E_e)\rangle &=&
\langle N^{(\rm SM)}(E_e)\rangle_{\rm SM} +
\Big\langle\frac{m_e}{E_e}\Big\rangle_{\rm SM}\frac{1}{1 +
  3\lambda^2}\,{\rm Re}(C_S C^*_T + \bar{C}_S \bar{C}^*_T + |C_T|^2 +
|\bar{C}_T|^2) + b_E = \nonumber\\ &=& 0.07767 + 0.6556\,\frac{1}{1 +
  3\lambda^2}\,{\rm Re}(C_S C^*_T + \bar{C}_S \bar{C}^*_T + |C_T|^2 +
|\bar{C}_T|^2) + b_E,\nonumber\\ \langle Q_{\rm e, eff}(E_e)\rangle
&=& \langle Q^{(\rm SM)}_e(E_e)\rangle_{\rm SM} + \frac{1}{1 +
  3\lambda^2}\,{\rm Re}\Big(C_S C^*_T + \bar{C}_S \bar{C}^*_T +
|C_T|^2 + |\bar{C}_T|^2\Big) - b_E = \nonumber\\ &=& 0.12279 +
\frac{1}{1 + 3\lambda^2}\,{\rm Re}\Big(C_S C^*_T + \bar{C}_S
\bar{C}^*_T + |C_T|^2 + |\bar{C}_T|^2\Big) - b_E,\nonumber\\ \langle
R_{\rm eff}(E_e)\rangle &=& \langle R^{(\rm SM)}(E_e)\rangle_{\rm SM}
+ \frac{1}{1 + 3\lambda^2}\,{\rm Im}(\lambda (C_S - \bar{C}_S) + (1 +
2 \lambda)(C_T - \bar{C}_T)) = \nonumber\\ &=&0.00089 + \frac{1}{1 +
  3\lambda^2}\,{\rm Im}(\lambda (C_S - \bar{C}_S) + (1 + 2
\lambda)(C_T - \bar{C}_T)).
\end{eqnarray}
For the calculation of $\langle X_{\rm eff}(E_e)\rangle$, where $X =
A_W, G, N, Q_e$ and $R$, we have used the electron--energy density
Eq.(\ref{eq:18}).

\subsection{Asymmetries of the neutron $\beta^-$--decay with polarized 
neutron and electron}

\subsubsection*{\bf Asymmetry of neutron--electron spin--momentum 
correlations: electron asymmetry}

For the electron asymmetry $A_{\exp}(E_e)$
\cite{Dubbers2008,Abele2008,Bopp1986}--\cite{Brown2018} we obtain the
following expression (see \cite{Ivanov2013})
\begin{eqnarray}\label{eq:22}
A_{\exp}(E_e) = \frac{{\cal N}^+_A(E_e) - {\cal N}^-_A(E_e)}{{\cal
    N}^+_A(E_e) + {\cal N}^-_A(E_e)} = \frac{1}{2}\,\beta\,{\cal A}_{W, \rm
  eff}(E_e) P_n (\cos\theta_1 + \cos\theta_2),
\end{eqnarray}
where $P_n = |\vec{\xi}_n| \le 1$ is the neutron spin polarization,
$\beta$ is the electron velocity and ${\cal N}^{\pm}_A(E_e)$ are the
numbers of events of the emission of the electron forward $(+)$ and
backward $(-)$ with respect to the neutron spin into the solid angle
$\Delta \Omega_{12} = 2\pi (\cos\theta_1 - \cos\theta_2)$ with $0 \le
\varphi \le 2\pi$ and $\theta_1 \le \theta_e \le \theta_2$. They are
determined by \cite{Dubbers2008} (see also \cite{Ivanov2013})
\begin{eqnarray}\label{eq:23}
\hspace{-0.3in}&&{\cal N}^{+}_A(E_e) = 2\pi {\cal N}(E_e)
\int^{\theta_2}_{\theta_1}\Big(1 + A_{W, \rm eff}(E_e)\,P_n\,
\beta\,\cos\theta_e\Big) \sin\theta_e\, d\theta_e =\nonumber\\
\hspace{-0.3in}&&= 2\pi {\cal N}(E_e)\,\Big(1 + \frac{1}{2}\,A_{W, \rm
  eff}(E_e)\,P_n\, \beta\,(\cos\theta_1 +
\cos\theta_2)\Big)\,(\cos\theta_1 - \cos\theta_2),\nonumber\\
\hspace{-0.3in}&&{\cal N}^{-}_A(E_e) = 2\pi {\cal N}(E_e)\int^{\pi -
  \theta_2}_{\pi - \theta_1} \Big(1 + \bar{A}_{W, \rm eff}(E_e)\,P_n\,
\beta\,\cos\theta_e\Big) \sin\theta_e\, d\theta_e = \nonumber\\
\hspace{-0.3in}&&= 2\pi {\cal N}(E_e)\,\Big(1 -
\frac{1}{2}\,\bar{A}_{W, \rm eff}(E_e)\,P_n\, \beta\,(\cos\theta_1 +
\cos\theta_2)\Big)\,(\cos\theta_1 - \cos\theta_2),
\end{eqnarray}
where ${\cal N}(E_e)$ is the normalization factor equal to
\begin{eqnarray}\label{eq:24}
\hspace{-0.3in}{\cal N}(E_e) &=& (1 + 3
\lambda^2)\,\frac{G^2_F|V_{ud}|^2}{8\pi^4} (E_0 - E_e)^2\,\sqrt{E^2_e
  - m^2_e}\,E_e\,F(E_e, Z = 1)\,\zeta^{(\rm SM)}(E_e)\nonumber\\
\hspace{-0.3in}&&\times\, \Big(1 + \frac{1}{2}\,\frac{1}{1 +
  3\lambda^2}\,( |C_S|^2 + |\bar{C}_S|^2 + 3 |C_T|^2 +
3|\bar{C}_T|^2)+ b_F\,\frac{m_e}{E_e}\Big).
\end{eqnarray}
The correlation coefficient ${\cal A}_{W, \rm eff}(E_e)$ in
Eq.(\ref{eq:22}) is given by
\begin{eqnarray}\label{eq:25}
{\cal A}_{W, \rm eff}(E_e) = \frac{\displaystyle A^{(\rm SM)}_W(E_e) -
  \frac{1}{1 + 3 \lambda^2}\,{\rm Re}( C_S \bar{C}^*_T + \bar{C}_S
  C^*_T + 2 C_T \bar{C}^*_T)}{\displaystyle 1 +
  \frac{1}{2}\,\frac{1}{1 + 3\lambda^2}\,( |C_S|^2 + |\bar{C}_S|^2 + 3
  |C_T|^2 + 3|\bar{C}_T|^2)+ b_F\,\frac{m_e}{E_e}} = \frac{A_{W,\rm
    eff}(E_e)}{\displaystyle 1 + b\,\frac{m_e}{E_e}}
\end{eqnarray}
with the correlation coefficient $A^{(\rm SM)}_W(E_e)$ equal to
\cite{Ivanov2013}
\begin{eqnarray}\label{eq:26}
\hspace{-0.3in}&&A^{(\rm SM)}_W(E_e) = \Big(1 +
\frac{\alpha}{\pi}\,f_n(E_e)\Big)\,A_0\Big\{1
-\frac{1}{M}\,\frac{1}{2\lambda (1 + \lambda)(1 + 3 \lambda^2)}\,
\Big(A^{(W)}_1 E_0 + A^{(W)}_2 E_e +
A^{(W)}_3\frac{m^2_e}{E_e}\Big)\Big\},\nonumber\\
\hspace{-0.3in}&&A^{(W)}_1 = \frac{2}{3}\,\Big(- 3 \lambda^3 + (3
\kappa + 5)\,\lambda^2 - (2 \kappa + 1)\,\lambda - (\kappa + 1)\Big) =
- 2\Big(\lambda - (\kappa + 1)\Big)\Big(\lambda^2 -
\frac{2}{3}\,\lambda - \frac{1}{3}\Big),\nonumber\\
\hspace{-0.3in}&&A^{(W)}_2 = \frac{2}{3}\,\Big(- 3 \lambda^4 + (3
\kappa + 12)\, \lambda^3 - (9 \kappa + 14)\,\lambda^2 + (5\kappa +
4)\,\lambda + (\kappa + 1)\Big) = - 2\Big(\lambda - (\kappa +
1)\Big)\, \Big(\lambda^3 - 3 \lambda^2 + \frac{5}{3}\,\lambda +
\frac{1}{3}\Big),\nonumber\\
\hspace{-0.3in}&&A^{(W)}_3 = - 4\,\lambda^2(\lambda +
1)\,\Big(\lambda - (\kappa + 1)\Big).
\end{eqnarray}
The electron asymmetry Eq.(\ref{eq:25}) can be used for the extraction
of contributions of interactions beyond the SM from the new
experimental data, which can be obtained in new runs of experiments on
the searches of interactions beyond the SM with cold and ultracold
neutrons \cite{Abele2016}.

\vspace{-0.1in}
\subsubsection*{\bf Asymmetry of electron spin--momentum correlations}

For the polarized electron and unpolarized neutron and proton the
correlation coefficient $G_{\rm eff}(E_e)$ defines the following
electron--energy and angular distribution
\begin{eqnarray}\label{eq:27}
\hspace{-0.3in}&&\frac{d^3 \lambda_n(E_e,\vec{k}_e, \vec{\xi}_e)}{dE_e
  d\Omega_e} = (1 + 3 \lambda^2)\,\frac{G^2_F|V_{ud}|^2}{8\pi^4}
\,(E_0 - E_e)^2 \sqrt{E^2_e - m^2_e}\, E_e\,F(E_e, Z = 1)\,\zeta^{(\rm
  SM)}(E_e)\nonumber\\
\hspace{-0.3in}&&\,\Big(1 + \frac{1}{2}\,\frac{1}{1 +
    3\lambda^2}\,( |C_S|^2 + |\bar{C}_S|^2 + 3 |C_T|^2 +
  3|\bar{C}_T|^2)\Big)\Big(1 + b\,\frac{m_e}{E_e} + G_{\rm
    eff}(E_e)\,\frac{\vec{\xi}_e \cdot \vec{k}_e}{E_e}\Big).
\end{eqnarray}
Following Kozela {\it et al.} \cite{Kozela2012} (see also
\cite{Ban2006}) the corresponding asymmetry can be defined as follows
\begin{eqnarray}\label{eq:28}
\hspace{-0.3in}&&G_{\exp}(E_e) = \frac{\displaystyle \frac{d^3
    \lambda_n(E_e,\vec{k}_e, \vec{\xi}_e)}{dE_e d\Omega_e}\Big|_+ -
  \frac{d^3 \lambda_n(E_e,\vec{k}_e, \vec{\xi}_e)}{dE_e
    d\Omega_e}\Big|_-}{\displaystyle \frac{d^3
    \lambda_n(E_e,\vec{k}_e, \vec{\xi}_e)}{dE_e d\Omega_e}\Big|_+ +
  \frac{d^3 \lambda_n(E_e,\vec{k}_e, \vec{\xi}_e)}{dE_e
    d\Omega_e}\Big|_-} = \beta\,{\cal G}_{\rm eff}(E_e)\,P_{e \parallel},
\end{eqnarray}
where $P_{e\parallel}$ is the longitudinal polarization of the
electron. The signs $(|_{\pm})$ mean parallel and anti--parallel
polarizations of the electron with respect to its momentum. For the
comparison with Ref.\cite{Kozela2012} we have to set $P_{e\parallel} =
\sigma_L$. The correlation coefficient ${\cal G}_{\rm eff}(E_e)$ is
equal to
\begin{eqnarray}\label{eq:29}
{\cal G}_{\rm eff}(E_e) &=& \frac{\displaystyle G^{(\rm SM)}(E_e) -
  \frac{1}{1 + 3\lambda^2}\,{\rm Re}\Big(C_S\bar{C}^*_S + 3
  C_T\bar{C}^*_T\Big)}{\displaystyle 1 + \frac{1}{2}\,\frac{1}{1 +
    3\lambda^2}\,( |C_S|^2 + |\bar{C}_S|^2 + 3 |C_T|^2 +
  3|\bar{C}_T|^2) + b_F\,\frac{m_e}{E_e}} = \frac{{G}_{\rm
    eff}(E_e)}{\displaystyle 1 + b\,\frac{m_e}{E_e}},
\end{eqnarray}
where $G^{(\rm SM)}(E_e)$ is given in Eq.(\ref{eq:5}).
\vspace{-0.1in}
\subsubsection*{\bf Asymmetry of neutron--electron spin--spin 
correlations}

For the decay electrons in the polarization states with polarization
$P_{e \perp}$ or $\sigma_{\rm T_1}$ in the notation of
Ref.\cite{Kozela2012}, lying in the decay plane spanned by the neutron
spin polarization vector $\vec{\xi}_n$ and electron momentum
$\vec{k}_e$ (see Fig.\,1 of Ref.\cite{Kozela2012}), we may define the
asymmetry \cite{Ban2006}, caused by the neutron--electron spin--spin
correlations
\begin{eqnarray}\label{eq:30}
\hspace{-0.3in}N_{\exp}(E_e) = {\cal N}_{\rm eff}(E_e)\,P_n \,P_{e
  \perp}\cos\gamma,
\end{eqnarray}
where we have denoted $\vec{\xi}_n\cdot \vec{\xi}_e = P_n \,P_{e
  \perp}\cos\gamma$.  The correlation coefficient ${\cal N}_{\rm
  eff}(E_e)$ in Eq.(\ref{eq:30}) is given by
\begin{eqnarray}\label{eq:31}
{\cal N}_{\rm eff}(E_e) = \frac{\displaystyle N^{(\rm
      SM)}(E_e) + \frac{m_e}{E_e}\, \frac{1}{1 + 3\lambda^2}\,{\rm
      Re}\Big(C_S C^*_T + \bar{C}_S \bar{C}^*_T + |C_T|^2 +
    |\bar{C}_T|^2\Big) + b_E}{\displaystyle 1 +
    \frac{1}{2}\,\frac{1}{1 + 3\lambda^2}\,( |C_S|^2 + |\bar{C}_S|^2 +
    3 |C_T|^2 + 3|\bar{C}_T|^2) + b_F\,\frac{m_e}{E_e}} =
  \frac{{N}_{\rm eff}(E_e)}{\displaystyle 1 + b\,\frac{m_e}{E_e}},
\end{eqnarray}
where $N^{(\rm SM)}(E_e)$ is defined in Eq.(\ref{eq:5}). The results,
obtained in this section, can be used for the analysis of the
experimental data of experiments on the searches of interactions
beyond the SM in the neutron $\beta^-$--decay with polarized neutron
and electron and unpolarized proton. The expressions
  for the correlations coefficients and asymmetries, obtained in this
  section, can be trivially defined in Gudkov's notation
  \cite{Gudkov2006} (see also \cite{Bhattacharya2012}) by a
  replacement $(C_j, \bar{C}_j) \to (C_j, C'_j)$ for $j = S,T$ (see
  Eq.(\ref{eq:8})).

\section{G--odd correlations}
\label{sec:gparity}

The $G$--parity transformation, i.e. $G = C\,e^{\,i \pi I_2}$, where
$C$ and $I_2$ are the charge conjugation and isospin operators, was
introduced by Lee and Yang \cite{Lee1956a} as a symmetry of strong
interactions. According to the $G$--transformation properties of
hadronic currents, Weinberg divided hadronic currents into two
classes, which are $G$--even first class and $G$--odd second class
currents \cite{Weinberg1958}, respectively. Thus, in agreement with
Weinberg's classification of hadronic currents the effective
phenomenological interactions beyond the SM Eq.(\ref{eq:6}) are
induced by the first class hadronic currents.

Following Weinberg \cite{Weinberg1958} and Gardner and Plaster
\cite{Gardner2013} the $G$--odd contributions or the contributions of
the second class hadronic currents to the matrix element of the
hadronic $n \to p$ transition in the $V - A$ theory of weak
interactions can be taken in the following form
\begin{eqnarray}\label{eq:32}
\langle
p(\vec{k}_p,\sigma_p)|J^{(+)}_{\mu}(0)|n(\vec{k}_n,\sigma_n)\rangle_{G
  - \rm odd} =
\bar{u}_p(\vec{k}_p,\sigma_p)\Big(\frac{q_{\mu}}{M}\,f_3(0) +
i\frac{1}{M}\,\sigma_{\mu\nu}\gamma^5 q^{\nu}g_2(0)\Big)\,u_n(\vec{k}_n,
\sigma_n),
\end{eqnarray}
where $J^{(+)}_{\mu}(0) = V^{(+)}_{\mu}(0) - A^{(+)}_{\mu}(0)$,
$\bar{u}_p(\vec{k}_p,\sigma_p)$ and $u_n(\vec{k}_n, \sigma_n)$ are the
Dirac wave functions of the proton and neutron \cite{Ivanov2018};
$f_3(0)$ and $g_2(0)$ are the phenomenological coupling constants
defining the strength of the second class currents in the weak decays.
The contribution of the second class currents Eq.(\ref{eq:32}) to the
amplitude of the neutron $\beta^-$--decay in the non--relativistic
baryon approximation is defined by
\begin{eqnarray}\label{eq:33}
 M(n \to p e^- \bar{\nu}_e)_{G - \rm odd} &=& -\,2m_n\,
\frac{G_F}{\sqrt{2}}\,V_{ud}\,\Big\{f_3(0)\,\frac{m_e}{M}\,
     [\varphi^{\dagger}_p\varphi_n] [\bar{u}_e (1 - \gamma^5)
       v_{\bar{\nu}}] +
     g_2(0)\,\frac{1}{M}\,[\varphi^{\dagger}_p(\vec{\sigma}\cdot
       \vec{k}_p)\varphi_n][\bar{u}_e \gamma^0 (1 - \gamma^5)
       v_{\bar{\nu}}]\nonumber\\
\hspace{-0.3in} && -
g_2(0)\,\frac{E_0}{M}\,[\varphi^{\dagger}_p\vec{\sigma}\varphi_n]
\cdot [\bar{u}_e \vec{\gamma}\,(1 - \gamma^5) v_{\bar{\nu}}] \Big\},
\end{eqnarray}
where we have kept only the leading $1/M$ terms in the large baryon
mass expansion.  The hermitian conjugate contribution is
\begin{eqnarray*}
M^{\dagger}(n \to p e^- \bar{\nu}_e)_{G - \rm odd} &=&
-\,2m_n\,
\frac{G_F}{\sqrt{2}}\,V_{ud}\,\Big\{f^*_3(0)\,\frac{m_e}{M}\,
     [\varphi^{\dagger}_n\varphi_p] [\bar{v}_{\nu} (1 + \gamma^5) u_e]
     + g^*_2(0)\,\frac{1}{M}\,[\varphi^{\dagger}_n(\vec{\sigma}\cdot
       \vec{k}_p)\varphi_p][\bar{v}_{\nu}\gamma^0 (1 - \gamma^5)
       u_e]\nonumber\\ 
\end{eqnarray*}
\begin{eqnarray}\label{eq:34}
&& -
     g^*_2(0)\,\frac{E_0}{M}\,[\varphi^{\dagger}_n\vec{\sigma}\varphi_p]
     \cdot [\bar{v}_{\nu} \vec{\gamma}\,(1 - \gamma^5) u_e] \Big\}.
\end{eqnarray}
The contributions of the $G$--odd correlations to the squared absolute
value of the amplitude of the neutron $\beta^-$--decay of polarized
neutron and electron and unpolarized proton, summed over polarizations
of massive fermions, are equal to
\begin{eqnarray}\label{eq:35}
 \hspace{-0.3in}&&\sum_{\rm pol.}\Big(M^{\dagger}(n \to p e^-
 \bar{\nu}_e)M (n \to p e^- \bar{\nu}_e)_{G - \rm odd} + M^{\dagger}
 (n \to p e^- \bar{\nu}_e)_{G - \rm odd}M(n \to p e^-
 \bar{\nu}_e)\Big) = 8m^2_n G^2_F|V_{ud}|^2\,\frac{1}{M} \nonumber\\
\hspace{-0.3in}&&\times \Big\{2\,{\rm Re}f_3(0)\,\frac{m^2_e}{E_e} +
2\,\lambda\,{\rm Re}f_3(0)\,m_e\Big(\vec{\xi}_n\cdot \vec{\xi}_e -
\frac{(\vec{\xi}_n\cdot \vec{k}_e)(\vec{k}_e\cdot \vec{\xi}_e)}{E_e
  (E_e + m_e)} \Big) + 2\,\lambda\,{\rm
  Im}f_3(0)\,m_e\,\frac{\vec{\xi}_n\cdot(\vec{k}_e\times
  \vec{\xi}_e)}{E_e}\nonumber\\
\hspace{-0.3in}&&+ 2\,{\rm Re}g_2(0)\,\Big[- \Big(\frac{4}{3}\,E_0 +
  \frac{2}{3}\,E_e\Big)\,\frac{\vec{\xi}_n\cdot \vec{k}_e}{E_e} +
  \Big(\frac{4}{3}\,E_0 -
  \frac{1}{3}\,E_e\Big)\,\frac{m_e}{E_e}\,(\vec{\xi}_n\cdot
  \vec{\xi}_e) + \Big(\frac{4}{3}\,E_0 + \frac{2}{3}\,E_e +
  m_e\Big)\,\frac{(\vec{\xi}_n\cdot\vec{k}_e)(\vec{k}_e\cdot
    \vec{\xi}_e)}{E_e(E_e + m_e)}\Big]\nonumber\\
\hspace{-0.3in}&& + 2\,\lambda\,{\rm Re}g_2(0)\,\Big[\Big(4 E_0 -
  \frac{m^2_e}{E_e}\Big) - 4\,E_0 \,\frac{\vec{\xi}_e\cdot
    \vec{k}_e}{E_e} + \Big(- \frac{8}{3}\,E_0 + \frac{2}{3}\,
  E_e\Big)\,\frac{\vec{\xi}_n\cdot \vec{k}_e}{E_e} +
  \Big(\frac{8}{3}\,E_0 -
  \frac{2}{3}\,E_e\Big)\,\frac{m_e}{E_e}\,(\vec{\xi}_n\cdot
  \vec{\xi}_e)\nonumber\\
\hspace{-0.3in}&&+ \Big(\frac{8}{3}\,E_0 -
\frac{2}{3}\,E_e\Big)\,\frac{(\vec{\xi}_n\cdot\vec{k}_e)(\vec{k}_e\cdot
  \vec{\xi}_e)}{E_e(E_e + m_e)}\Big] - 2\,\lambda\,{\rm Im}g_2(0)\,
m_e\,\frac{\vec{\xi}_n\cdot(\vec{k}_e\times \vec{\xi}_e)}{E_e}\Big\}.
\end{eqnarray}
For the relative $G$--odd contributions to the correlation
coefficients we obtain the following expressions
\begin{eqnarray}\label{eq:36}
 \hspace{-0.3in}\frac{\delta \zeta(E_e)_{G-\rm odd}}{\zeta^{(\rm
     SM)}(E_e)} &=& \frac{2}{1 + 3\lambda^2}\,\frac{1}{M}\,\Big\{{\rm
   Re}f_3(0)\,\frac{m^2_e}{E_e} + \lambda\,{\rm Re}g_2(0)\,\Big(4 E_0
 - \frac{m^2_e}{E_e}\Big)\Big\},\nonumber\\
\hspace{-0.3in}\frac{\delta A_W(E_e)_{G-\rm odd}}{ A^{(\rm
    SM)}_W(E_e)} &=& \frac{2}{1 + 3\lambda^2}\,\frac{1}{M}\,\frac{{\rm
    Re}g_2(0)}{A_0}\,\Big\{\Big(- \frac{8}{3}\,\lambda -
\frac{4}{3}\Big)\,E_0 + \Big(\frac{2}{3}\,\lambda -
\frac{2}{3}\Big)\,E_e\Big\} - \delta \zeta(E_e)_{G-\rm odd},\nonumber\\
\hspace{-0.3in}\frac{\delta G(E_e)_{G-\rm odd}}{G^{(\rm SM)}(E_e)} &=&
\frac{2\lambda}{1 + 3\lambda^2}\,\frac{4E_0}{M}\,{\rm Re}g_2(0) -
\delta \zeta(E_e)_{G-\rm odd},\nonumber\\
\hspace{-0.3in}\frac{\delta N(E_e)_{G-\rm odd}}{N^{(\rm SM)}(E_e)} &=&
- \frac{2}{1 + 3\lambda^2}\,\frac{1}{M}\,\frac{1}{A_0}\,
\Big\{\lambda\,{\rm Re}f_3(0)\,E_e + {\rm
  Re}g_2(0)\,\Big[\Big(\frac{8}{3}\,\lambda + \frac{4}{3}\Big)\,E_0 -
  \Big(\frac{2}{3}\,\lambda + \frac{1}{3}\Big)\,E_e
  \Big]\Big\}- \delta \zeta(E_e)_{G-\rm odd},\nonumber\\
\hspace{-0.3in}\frac{\delta Q_e(E_e)_{G-\rm odd}}{Q^{(\rm SM)}_e(E_e)}
&=& - \frac{2}{1 + 3\lambda^2}\,\frac{1}{M}\,\frac{1}{A_0}\, \Big\{ -
\lambda\,{\rm Re}f_3(0)\,m_e + {\rm
  Re}g_2(0)\,\Big[\Big(\frac{8}{3}\,\lambda + \frac{2}{3}\Big)\,E_0 -
  \Big(\frac{2}{3}\,\lambda - \frac{2}{3}\Big)\,E_e + m_e\Big]\Big\}
\nonumber\\
\hspace{-0.3in}&& -
\delta \zeta(E_e)_{G-\rm odd},\nonumber\\
\hspace{-0.3in}\frac{\delta R(E_e)_{G-\rm odd}}{R^{(\rm SM)}_e(E_e)}
&=& - \frac{2\lambda}{1 + 3\lambda^2}\,\frac{k_e}{M}\,\frac{1}{\alpha
  A_0}\,{\rm Im}\big(f_3(0) - g_2(0)\big).
\end{eqnarray}
For $\lambda = - 1.2750$ \cite{Abele2008} we get
\begin{eqnarray}\label{eq:37}
 \hspace{-0.3in}\frac{\delta \zeta(E_e)_{G-\rm odd}}{\zeta^{(\rm
     SM)}(E_e)} &=& 1.85\times 10^{-4}\,{\rm
   Re}f_3(0)\,\frac{m_e}{E_e} + \Big(- 2.39\times 10^{-3} + 2.36\times
 10^{-4}\,\frac{m_e}{E_e}\Big)\,{\rm Re}g_2(0), \nonumber\\
\hspace{-0.3in}\frac{\delta A_W(E_e)_{G-\rm odd}}{ A^{(\rm
    SM)}_W(E_e)} &=& - 1.85\times 10^{-4}\,{\rm
  Re}f_3(0)\,\frac{m_e}{E_e} + \Big(- 5.73\times 10^{-3} + 5.96\times
10^{-3}\,\frac{E_e}{E_0} - 2.36\times
10^{-4}\,\frac{m_e}{E_e}\Big)\,{\rm Re}g_2(0),\nonumber\\
\hspace{-0.3in}\frac{\delta G(E_e)_{G-\rm odd}}{G^{(\rm SM)}(E_e)} &=&
- 1.85\times 10^{-4}\,{\rm Re}f_3(0)\,\frac{m_e}{E_e} - 2.36\times
10^{-4}\,{\rm Re}g_2(0)\,\frac{m_e}{E_e},\nonumber\\
\hspace{-0.3in}\frac{\delta N(E_e)_{G-\rm odd}}{N^{(\rm SM)}(E_e)} &=&
\Big(- 5.00\times 10^{-3}\,\frac{E_e}{E_0} - 1.85\times
10^{-4}\,\frac{m_e}{E_e}\Big)\,{\rm Re} f_3(0) + \Big(- 5.73\times
10^{-3} + 2.03\times 10^{-3}\,\frac{E_e}{E_0} \nonumber\\
\hspace{-0.3in}&&- 2.36\times
10^{-4}\,\frac{m_e}{E_e}\Big)\,{\rm Re} g_2(0),\nonumber\\
\hspace{-0.3in}\frac{\delta Q_e(E_e)_{G-\rm odd}}{Q^{(\rm SM)}_e(E_e)}
&=& \Big(1.98\times 10^{-3} - 1.85\times
10^{-4}\,\frac{m_e}{E_e}\Big)\,{\rm Re} f_3(0) + \Big(- 6.79\times
10^{-3} + 5.96\times 10^{-3}\,\frac{E_e}{E_0}\nonumber\\
\hspace{-0.3in}&& - 2.36\times
10^{-4}\,\frac{m_e}{E_e}\Big){\rm Re} g_2(0),\nonumber\\
\hspace{-0.3in}\frac{\delta R(E_e)_{G-\rm odd}}{R^{(\rm SM)}_e(E_e)}
&=& - 0.69\,\frac{k_e}{E_0}\,{\rm Im}\big(f_3(0) - g_2(0)\big).
\end{eqnarray}
The $G$--odd correction to the neutron lifetime is
\begin{eqnarray}\label{eq:38}
\frac{1}{\tau^{(\rm eff)}_n} &=& \frac{1}{\tau_n}\Big\{1 + \frac{2}{1
  + 3\lambda^2}\,\frac{1}{M}\,\Big[\Big({\rm
    Re}f_3(0)\,\Big\langle\frac{m^2_e}{E_e}\Big\rangle_{\rm SM} +
  \lambda\,{\rm Re}g_2(0)\,\Big(4 E_0 -
  \Big\langle\frac{m^2_e}{E_e}\Big\rangle_{\rm SM}\Big)\Big]\Big\}
=\nonumber\\ &=& \frac{1}{\tau_n}\,\big(1 + 1.21\times 10^{-4}\,{\rm
  Re}f_3(0) - 22.35\times 10^{-4}\,{\rm Re}g_2(0)\big),
\end{eqnarray}
where $\langle m_e/E_e\rangle_{\rm SM} = 0.6556$. For $|{\rm
  Re}f_3(0)| < 0.1$ and $|{\rm Re}g_2(0)| < 0.01$ the contributions of
the $G$--odd correlations to the neutron lifetime and correlation
coefficients can appear at the level of $10^{-5}$ or even
smaller. This agrees well with results obtained by Gardner and Plaster
\cite{Gardner2013}.

\section{Conclusion}
\label{sec:conclusion}

In this paper we have continued our work on the precision analysis of
the neutron lifetime and the correlation coefficients of the
electron--energy and angular distribution of the neutron
$\beta^-$--decay with polarized neutron and electron and unpolarized
proton. The correlation coefficients, calculated within the SM with
Wilkinson's corrections \cite{Ivanov2017b}, we have supplemented by
the contributions of interactions beyond the SM. Since the
contributions of vector and axial vector interactions beyond the SM,
calculated to linear approximation, can be absorbed by the axial
coupling constant $\lambda$ and the CKM matrix element $V_{ud}$
\cite{Ivanov2013}, the observable contributions of interactions beyond
the SM are defined by scalar and tensor interactions only. We have
taken into account a complete set of contributions of scalar and
tensor interactions, which have been calculated in the linear
approximation for the vector and axial--vector interactions beyond the
SM \cite{Ivanov2013} and to leading order in the large nucleon mass
expansion.  The neutron lifetime and asymmetries of the neutron
$\beta^-$--decay with polarized neutron and electron, calculated in
section \ref{sec:asymmetry}, can be used for the analysis of
experimental data of experiments on the searches of contributions of
interactions beyond the SM.

\begin{figure}
\includegraphics[height=0.20\textheight]{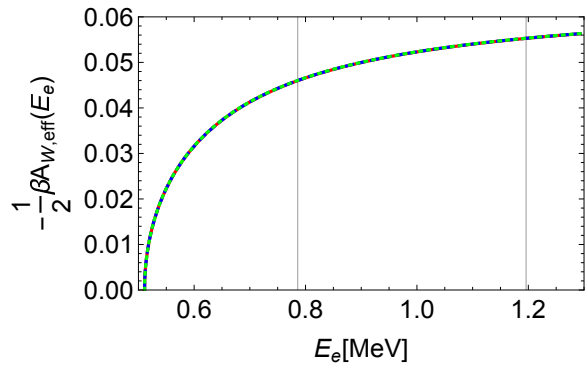}
  \caption{The theoretical electron asymmetry, calculated in the SM
    (blue dotted curve) and with interactions beyond the SM (red
    dotted curve) for  $b_F = - 0.0002$
    and (green dotted curve) for $b_F = - 0.0004$, respectively. }
\label{fig:fig1}
\end{figure}

\vspace{-0.1in}

\subsubsection*{\bf Numerical estimates of contributions of interactions 
beyond the SM}

\vspace{-0.1in}

Using the results, obtained in this paper, we may attempt to make some
estimates of the values of scalar and tensor coupling constants and
contributions of scalar and tensor interactions to
observables. Following \cite{Ivanov2013} and \cite{Abele2016} we
assume that the contributions of scalar and tensor interactions beyond
the SM to neutron lifetime are at the level $10^{-4}$.  As we have
shown (see Eq.(\ref{eq:20})), the real parts of the scalar and tensor
coupling constants are defined in terms of the correlation
coefficients $b_F$ and $b_E$.  According to recent analysis by Hardy
and Towner \cite{Hardy2015}, the value of the Fierz interference term
is equal to $b = - 0.0028 \pm 0.0026$. This allows to analyse the
values of the Fierz interference term from the interval $- 0.0054 \le
b \le - 0.0002$. Of course, we understand that the estimate of the
Fierz interference term $b = - 0.0028 \pm 0.0026$ is obtained from the
pure Fermi $0^+ \to 0^+$ transitions, caused by the vector part of the
effective $V - A$ weak interactions. Thus, according to the definition
of the Fierz interference term (see Eq.(\ref{eq:14}) and
Eq.(\ref{eq:15})), it should be induced only by scalar interactions
beyond the SM. Nevertheless, in spite of this fact we propose to use
such an estimate $b = - 0.0028 \pm 0.0026$ in more extended
interpretation, allowing to understand the order of contributions of
interactions beyond the SM to the neutron lifetime and correlation
coefficients of the neutron $\beta^-$--decays at the level of
$10^{-4}$.

For an estimate of the strengths of the scalar and tensor interactions
beyond the SM we accept for simplicity the approximation by real
scalar and tensor coupling constants and nucleon--lepton four--fermion
interactions with the left--handed neutrinos only
\cite{Severijns2006}, i.e. $C_S = - \bar{C}_S$ and $C_T = -
\bar{C}_T$. In order to fit the mean value of the experimental data on
the averaged value of the correlation coefficient of the
neutron--electron spin--spin correlations $\langle N_{\rm
  eff}(E_e)\rangle = 0.0670$ we have to set $b_E = - 0.0107$. In Table
I we give the values of the scalar and tensor coupling constants and
their contributions to the neutron lifetime and measurable correlation
coefficients.

Below without loss of generality instead of the Fierz interferance
term $b$ we use the correlation coefficients $b_F$. From Table I one
may see that the correlation coefficient $b_F$ coincides with the
Fierz interference term with an accuracy of about $7\times 10^{-5}$.
Setting $b_F = - 0.0054$ we find that the contribution of interactions
beyond the SM to the neutron lifetime is at the level of $3\sigma$
with respect to the world averaged value $\tau_n = 880.2(1.0)\,{\rm
  s}$ \cite{PDG2016} and the experimental one $\tau_n =
880.2(1.2)\,{\rm s}$ \cite{Arzumanov2015}. This provides sufficiently
strong deviation of the theoretical value of the neutron lifetime of
about $3\,{\rm s}$ from the experimental one $\tau_n =
880.2(1.2)\,{\rm s}$ \cite{Arzumanov2015} and something of about
$5\,{\rm s}$ from recent experimental value $\tau_n = 877.7 \pm
0.7_{\rm stat.}{^{+0.3}_{-0.1}}_{\rm syst.}\,{\rm s}$, which has been
reported by the UCNA Collaboration \cite{Pattie2017}. If we want to
keep the contribution of interactions beyond the SM to the neutron
lifetime at the level of $1\sigma$ or $10^{-3}$ we have to restrict
the values of the correlation coefficients $b_F$ as follows $- 0.0017
\le b_F \le - 0.0002$. However, in order to keep the contribution of
interactions beyond the SM to the neutron lifetime at the level of
$10^{-4}$ or even smaller we have to analyse the contributions of the
correlation coefficient $b_F$ with the values taken from the interval
$- 0.0004 \le b_F \le - 0.0002$.

In Fig.\,\ref{fig:fig1} we plot the theoretical electron asymmetry,
calculated within the SM (blue dotted line) with $A_{W, \rm eff} (E_e)
= A^{(\rm SM)}_W(E_e)$ (see Eq.(\ref{eq:26})) only and within the SM
with the contributions of interactions beyond the SM, given by
Eq.(\ref{eq:25}) and calculated for the correlation coefficient $b_F =
- 0.0002$ (red dotted line) and $b_F = - 0.0004$ (green dotted
line). The vertical lines in Fig.\,\ref{fig:fig1} define the
experimental electron--energy region of the electron asymmetry
observation. We would like to emphasize that obtained estimates of
contributions of scalar and tensor interactions beyond the SM to the
electron--asymmetry do not contradict the experimental data on the
correlation coefficient $A_0$ and the axial coupling constant
$\lambda$, extracted from experimental data of recent measurements of
the electron--asymmetry, namely: $A_0 = - 0.11933(34)$ ($\lambda = -
0.12750(9)$ \cite{Abele2008}, $A_0 = - 0.11966(89)(^{+123}_{-140})$
($\lambda = - 1.27590(239)(^{+331}_{-377})$) \cite{Plaster2012}, $A_0
= - 0.11972(45)_{\rm stat.}(^{+32}_{-44})_{\rm syst.}$ ($\lambda = -
1.2761(12)_{\rm stat.}(^{+9}_{-12})_{\rm syst.}$) \cite{Mund2013},
$A_0 = - 0.11954(55)_{\rm stat.}(98)_{\rm syst.}$ ($\lambda = -
1.2756(30)$) \cite{Mendenhall2013}, and $A_0 = - 0.12015(34)_{\rm
  stat.}(63)_{\rm syst.}$ ($\lambda = - 1.2772(20)$) and $A_0 = -
0.12054(44)_{\rm stat.}(68)_{\rm syst.}$ ($\lambda = -
1.2783(22)$)\cite{Brown2018}. Varying the axial coupling constant from
$\lambda = - 1.2750$ \cite{Abele2008} to $\lambda = - 1.2783$
\cite{Brown2018} and keeping $b_F = - 0.0002$ and $b_E = - 0.0107$ one
may show that the scalar and tensor couping constants change their
values by $\Delta C_S = 3.56\times 10^{-5}$ and $\Delta C_T = -
2.99\times 10^{-6}$, respectively.

It is important to emphasize that the contributions of scalar and
tensor interactions beyond the SM to the correlation coefficients
$\bar{N}_{\rm eff}(E_e)$ and $N_{\rm eff}(E_e)$, caused by
neutron--electron spin--spin correlations, are of order $10^{-2}$ and
do not practically depend on the values of the correlation coefficient
$b_F$ (or the Fierz interference term $b$ at the level of accuracy of
about $7\times 10^{-5}$) taken from the interval $- 0.0054 \le b_F \le
- 0.0002$. The contributions of interactions beyond the SM to these
correlation coefficients are practically defined by the correlation
coefficient $b_E$, which we have set equal to $b_E = - 0.0107$. Of
course, our estimate depends strongly on the experimental mean--value
$N_{\exp} = 0.067 \pm 0.011_{\rm stat.}\pm 0.004_{\rm syst.}$ of the
correlation coefficient of the neutron--electron spin--spin
correlations, which we have accepted as a signal for a trace of
contributions of interactions beyond the SM. Of course, such an
assumption seems to be sufficiently strong if to take into account
that the experimental value $N_{\exp} = 0.067 \pm 0.011_{\rm stat.}\pm
0.004_{\rm syst.}$ agrees with the theoretical one $\langle N^{(\rm
  SM)}(E_e)\rangle_{\rm SM} = 0.07767$, calculated in the SM
\cite{Ivanov2017b}, within one standard deviation.

Of course, the contributions of interactions beyond the SM of order
$10^{-2}$ to the correlation coefficients $N_{\rm eff}(E_e)$, $Q_{e
  \rm eff}(E_e)$ and $\bar{N}_{\rm eff}(E_e)$ seem to be unreal, and
we have to keep them at the level of $10^{-4}$.  In Table II we give
some estimates of the scalar and tensor coupling constants obtained
for the correlation coefficients $b_F = - 0.0002$ and $|b_E| \sim
10^{-4}$.

\begin{table}[h]
\begin{tabular}{|c|c|c|c|c|c|c|c|}
\hline $b_F$ & $b_E$ & $C_S$ & $C_T$ & $ \Delta \tau^{(\rm
  BSM)}_n/\tau_n$ & $\langle \bar{N}^{(\rm BSM)}(E_e)\rangle_{\rm SM}$
& $\langle N^{(\rm BSM)}(E_e)\rangle_{\rm SM} $ & $\langle A^{(\rm
  BSM)}_W(E_e)\rangle_{\rm SM} $\\ \hline $~- 0.0002~$ & $~- 0.0107~$
& $~0.0186~$ & $~0.0050~$ & $6.0\times 10^{-5}$ & $- 0.00944$ & $-
0.01067$ & $4.0\times 10^{-5}$\\\hline $~- 0.0054~$ & $~- 0.0107~$ &
$~0.0149~$ & $~0.0080~$ & $3.5\times 10^{-3}$ & $- 0.00942$ & $-
0.01066$ & $6.3\times 10^{-5}$\\ \hline $~- 0.0017~$ & $~ - 0.0107~$ &
$~0.0175~$ & $~0.0059~$ & $10^{-3}$ & $- 0.00944$ & $- 0.01067$ &
$4.7\times 10^{-5}$\\\hline $~- 0.0004~$ & $~ - 0.0107~$ & $~0.0184~$
& $~0.0051~$ & $1.9\times 10^{-4}$ & $- 0.00944$ & $- 0.01067$ &
$4.1\times 10^{-5}$\\\hline
\end{tabular} 
\caption{Scalar and tensor coupling constants of interactions beyond
  the SM and their contributions to the neutron lifetime and the
  measurable correlation coefficients of the neutron $\beta^-$--decay
  with polarized neutron and electron for the correlation coefficient
  $b_E = - 0.0107$. With an accuracy of about $7\times 10^{-5}$ the
  values of the correlation coefficient $b_F$ define the values of the
  Fierz interference term $b$.}
\end{table}
\begin{table}[h]
\begin{tabular}{|c|c|c|c|c|c|c|c|}
\hline $b_F$ & $b_E$ & $C_S$ & $C_T$ & $ \Delta \tau^{(\rm
  BSM)}_n/\tau_n$ & $\langle \bar{N}^{(\rm BSM)}(E_e)\rangle_{\rm SM}$
& $\langle N^{(\rm BSM)}(E_e)\rangle_{\rm SM} $ & $\langle A^{(\rm
  BSM)}_W(E_e)\rangle_{\rm SM} $\\ \hline $~- 0.0002~$ & $~- 0.0002~$
& $+ 2.1\times 10^{-4}$ & $+ 2.1\times 10^{-4}$ & $1.3\times 10^{-4}$
& $- 1.77\times 10^{-4}$ & $- 2.00 \times 10^{-4}$ & $+ 2.95\times
10^{-8}$\\\hline $~- 0.0002~$ & $~+ 0.0002~$ & $- 4.9\times 10^{-4}$ &
$+ 2.5\times 10^{-5}$ & $1.3\times 10^{-4}$ & $+ 1.77\times 10^{-4}$ &
$+ 2.00\times 10^{-4}$ & $- 3.99 \times 10^{-9}$\\ \hline $~- 0.0002~$
& $ ~- 0.0001~$ & $+ 3.3\times 10^{-5}$ & $+ 1.6\times 10^{-4}$ &
$1.3\times 10^{-4}$ & $- 8.85\times 10^{-5}$ & $- 1.00\times 10^{-4}$
& $+ 1.08\times 10^{-8}$\\\hline $~- 0.0002~$ & $~ + 0.0001~$ & $-
3.2\times 10^{-4}$ & $+ 7.1\times 10^{-5}$ & $1.3\times 10^{-4}$ & $+
8.85\times 10^{-5}$ & $+ 1.00\times 10^{-4}$ & $- 5.93\times
10^{-9}$\\\hline
\end{tabular} 
\caption{Scalar and tensor coupling constants of interactions beyond
  the SM and their contributions to the neutron lifetime and the
  measurable correlation coefficients of the neutron $\beta^-$--decay
  with polarized neutron and electron for the correlation coefficient
  $b_E$ taken at the level of $10^{-4}$.}
\end{table}

It is interesting that for $b_F = - 0.0002$ and keeping the value of
the correlation coefficient $b_E$ at the level of $10^{-4}$,
i.e. $|b_F| \sim |b_E| \sim 10^{-4}$, we get the results, which are
adduced in Table II.  One may see that for the correlation
coefficients $b_F$ and $b_E$ kept at the level of $10^{-4}$ the values
of scalar and tensor coupling constants are at the level of $10^{-5} -
10^{-4}$. In this case the contributions of scalar and tensor
interactions beyond the SM can be taken in the linear approximation
and fully defined by $b_F$ and $b_E$. As a result, expected
experimental mean values of the correlation coefficient $\langle
N_{\rm eff}(E_e)\rangle$ of the neutron--electron spin--spin
correlations, averaged over the electron--energy spectrum, may appear,
for example, from the interval $0.07747 \le \langle N_{\rm
  eff}(E_e)\rangle \le 0.07787$.

\subsubsection*{\bf Towards robust SM theoretical background 
with corrections to order $10^{-5}$ for analysis of experimental data
of experiments on searches of interactions beyond the SM at the level
of $10^{-4}$}

It is obvious that the analysis of experimental data of experiments on
the searches of contributions of interactions beyond the SM at the
level of $10^{-4}$ or even better \cite{Abele2016} demands a robust SM
theoretical background with corrections at the level of
$10^{-5}$. These are i) Wilkinson's corrections \cite{Ivanov2017b} and
ii) corrections of order $O(E^2_e/M^2)$ defined by the weak magnetism
and proton recoil, calculated to next--to--next--to--leading order in
the large nucleon mass expansion, the radiative corrections of order
$O(\alpha E_e/M)$, calculated to next--to--leading order in the large
nucleon mass expansion, and the radiative corrections of order
$O(\alpha^2/\pi^2)$, calculated to leading order in the large nucleon
mass expansion \cite{Ivanov2017c}.  These theoretical corrections
should provide for the analysis of experimental data of "discovery"
experiments the required $5\sigma$ level of experimental uncertainties
of a few parts in $10^{-5}$ \cite{Ivanov2017b}. An important role of
strong low--energy interactions for a correct gauge invariant
calculation of radiative corrections of order $O(\alpha E_e/M)$ and
$O(\alpha^2/\pi^2)$ as functions of the electron energy $E_e$ has been
pointed out in \cite{Ivanov2017c}. This agrees with Weinberg's
assertion about important role of strong low--energy interactions in
decay processes \cite{Weinberg1957}. A procedure for the calculation
of these radiative corrections to the neutron $\beta^-$--decays with a
consistent account for contributions of strong low--energy
interactions, leading to gauge invariant observable expressions
dependent on the electron energy $E_e$ determined at the confidence
level of Sirlin's radiative corrections \cite{Sirlin1967}, has been
proposed in \cite{Ivanov2017c}.

The contributions of the $G$--odd correlations or the contributions of
the second class hadronic currents \cite{Weinberg1958} we have found
at the level of $10^{-5}$ or even smaller. Such an estimate does not
contradict the estimates performed by Gardner and Plaster
\cite{Gardner2013}. It is just the level of the SM corrections by
Wilkinson \cite{Wilkinson1982} and corrections of order $O(\alpha
E_e/M)$, $O(\alpha^2/\pi^2)$ and $O(E^2_e/M^2)$ pointed out in
\cite{Ivanov2017c}. These SM corrections should be taken into account
for experimental searches for interactions beyond the SM of order
$10^{-4}$, caused by the contributions of the first class hadronic
currents (see Eq.(\ref{eq:6})) \cite{Weinberg1958}, where a
"discovery" experiment with the required 5$\sigma$ sensitivity will
require experimental uncertainties of a few parts in $10^{-5}$
\cite{Ivanov2017b}. An estimate of the $G$--odd correlations or
contributions of the second class hadronic currents at the level of
$10^{-5}$ implies an urgent necessity of the robust theoretical
background, caused by the Wilkinson corrections and corrections of
order $O(\alpha E_e/M)$, $O(\alpha^2/\pi^2)$ and $O(E^2_e/M^2)$. A
specific dependence of the $G$--odd corrections on the electron energy
should allow to distinguish them from the SM background corrections of
order $10^{-5}$ and the contributions of interactions beyond the SM of
order $10^{-4}$, caused by the first class hadronic currents. Thus,
one may argue that just after the calculation of the theoretical
background of order $10^{-5}$, caused by the Wilkinson corrections and
corrections $O(\alpha E_e/M)$, $O(\alpha^2/\pi^2)$ and $O(E^2_e/M^2)$,
a perspective of an experimental discovery of the contributions of the
second class hadronic currents (or the $G$--odd corrections) as well
as the contributions of the first class hadronic currents beyond the
SM, should not be illusive and unfeasible.


\section{Acknowledgements}

We thank Hartmut Abele for fruitful discussions and comments. The work
of A. N. Ivanov was supported by the Austrian ``Fonds zur F\"orderung
der Wissenschaftlichen Forschung'' (FWF) under contracts P26781-N20
and P26636-N20 and ``Deutsche F\"orderungsgemeinschaft'' (DFG) AB
128/5-2. The work of R. H\"ollwieser was supported by the Deutsche
Forschungsgemeinschaft in the SFB/TR 55. The work of M. Wellenzohn was
supported by the MA 23 (FH-Call 16) under the project ``Photonik -
Stiftungsprofessur f\"ur Lehre''.


\begin{thebibliography}{9}
\bibitem{SPT1a} 
There is also the pseudoscalar lepton--baryon weak
  coupling constant, the contribution of which can be neglected with
  respect to the contributions of order $10^{-5}$.
\end{thebibliography}
\end{document}